\documentclass[%
 reprint,
aps,
prd
]{revtex4-2}

\usepackage{graphicx,amsmath,amssymb,xcolor}
\newcommand{\de}{\text{d}}
\newcommand{\D}{\text{D}}

\begin{document}

\title{The First Law of Black Hole Thermodynamics in Arbitrary Gravity and Matter Theories Using Differential Forms}

\author{O. Ram\'irez}

\author{Y. Bonder}
\email{bonder@nucleares.unam.mx}
\affiliation{Instituto de Ciencias Nucleares, 
Universidad Nacional Aut\'onoma de M\'exico\\
Apartado Postal 70-543, Ciudad de M\'exico, 04510, M\'exico}

\begin{abstract}
Black hole thermodynamics plays a central role in probing the microscopic structure of spacetime. Since the 1990s, a general method for deriving the first law of black hole thermodynamics in arbitrary diffeomorphism-invariant theories has been available within the metric formalism. Differential forms, however, provide a natural and powerful language for formulating gravitational theories, particularly those extending beyond General Relativity. In this work, we develop a unified prescription for deriving the first law using differential forms defined on spacetime. Our framework applies to a broad class of theories, including those with torsion and generic matter fields, and yields concrete expressions for the first law.
\end{abstract}

\maketitle
\section{Introduction}

Black hole thermodynamics was pioneered by Bekenstein~\cite{Bekenstein1972,Bekenstein1973}, who uncovered a remarkable analogy between expressions governing stationary black holes and the laws of thermodynamics. As such, this correspondence provides one of the few available windows into the underlying gravitational microstates.

In this analogy, the area of a black hole horizon, shown by Hawking to be nondecreasing~\cite{PhysRevLett.26.1344}, plays the role of entropy. The correspondence is further strengthened by another of Hawking’s discoveries: that black holes emit thermal radiation due to quantum effects~\cite{Hawking1975}, thereby establishing that the quantity identified as temperature in the analogy corresponds to a genuine physical temperature.

Black hole thermodynamics was originally derived within General Relativity (GR), the most successful theory of gravitation to date~\cite{Will}. However, it is expected to persist in more refined theories of gravity~\cite{Clifton2012}. Recall that, if gravity admits a quantum description, GR is expected to only be the leading contribution of a more fundamental framework. Hence, establishing a connection between black hole thermodynamics and such generalized theories is of central importance.

A natural language for formulating gravitational theories is provided by the formalism of differential forms~\cite{BlagojevicHehl2013,Ortin2015} in which the action is expressed as the integral of a Lagrangian top-form constructed from the relevant dynamical fields. An efficient realization of this framework is the first-order formalism of gravity~\cite{Cartan1981}, in which the vielbein and the spin connection are the independent gravitational variables. Importantly, an independent spin connection usually gives rise to torsion, which is yet to be empirically falsified~\cite{YuriBruno}. 

Wald and collaborators developed a general method to compute black hole entropy from the Noether charge associated with diffeomorphism invariance~\cite{wald,Iyer}. This method applies to a broad class of diffeomorphism-invariant gravitational theories formulated within the metric formalism. This approach can be extended to the language of differential forms using the framework of principal fiber bundles~\cite{Prabhu_2017}. In the context of principal fiber bundles, gauge independence was established by defining horizontal differential forms, such as the Lagrangian. Since horizontal forms are insensitive to the vertical directions of the bundle, which are associated with gauge transformations, the resulting formulation is gauge independent. Hence, it provides a rigorous geometric treatment of gauge symmetries and yields a gauge-independent formulation of black hole thermodynamics.

The method presented here achieves the same goal while working entirely with fields defined on spacetime. In particular, it avoids the intermediate steps associated with principal bundle constructions, such as introducing horizontal lifts, performing calculations on the total space, and projecting the resulting quantities back to spacetime. As a result, the derivation remains manifestly covariant and gauge independent, while being technically simpler. Thus, it enables the derivation of explicit expressions within a unified framework. As a result, a concrete formula is obtained that generalizes earlier results, many of which were derived on a case-by-case basis within specific theories (see, e.g., Refs.~\onlinecite{Blagojevic2019_EntropyPoincareHamiltonian,Blagojevic2020_Entropy3DGRKerrAdS,Blagojevic2020_EntropyGRKerrAdS,Blagojevic2020_EntropyPoincareKerrAdS,Blagojevic2021_ThermodynamicsKerrAdSPGTheory,Blagojevic2021_ThermodynamicsRiemannianKerrAdS,Blagojevic2022_EntropyKerrNewmanAdSTorsion,Blagojevic2022_EntropyReissnerNordstromLike,Blagojevic2023_EntropyCoupledScalar,Blagojevic2023_ThermodynamicsChargedScalar,Ballesteros_2024}).

This paper is organized as follows. Section~\ref{sec:preliminaries} reviews the necessary background, including the relation between the metric and first-order formalisms, together with a brief account of the previous method for deriving black hole entropy. In Sec.~\ref{BHEFOF}, the method is generalized to the first-order formalism, and the relevant gauge-invariant quantities entering the first law are identified. Section~\ref{SecGTGl} considers a general diffeomorphism-invariant gravitational theory with specified dependence on torsion and curvature, for which the quantities entering the first law can be computed explicitly. The inclusion of matter fields is discussed in Sec.~\ref{MFie}, and the entropy expression is derived in Sec.~\ref{Ent}. Aspects of the Hamiltonian formulation in the language of differential forms are analyzed in Sec.~\ref{HamInv}. The conclusions are summarized in Sec.~\ref{sec:conclusions}. Additional technical details are provided in the appendices.

The following notation is adopted throughout this paper. Latin lowercase indices from the beginning of the alphabet denote abstract spacetime indices and are omitted when working with differential forms. Greek indices label internal group indices (typically associated with the Lorentz group, though they may be associated with an arbitrary Lie group in more general theories), while capital Latin indices are reserved for Lie-algebra indices associated with matter fields. Index raising and lowering are performed using the appropriate metric and its inverse, which are distinguished by the position of their indices. For example, spacetime indices are raised and lowered with the metric \( g_{ab} \), while internal indices are handled using the metric associated with the relevant group, denoted by \( \eta_{\mu\nu} \). A sum over its entire range is understood for repeated indices. Conventions for operations on differential forms follow Refs.~\onlinecite{review,Naka}. All derivatives act on the object that directly follows them, while the integral sign applies to the entire expression. When no integration domain is specified, it is understood to be an arbitrary region of spacetime. All fields are assumed to be smooth.

\section{Preliminaries}\label{sec:preliminaries}

To make the paper self-contained, the essential background material is reviewed here. This section is divided into two subsections: the first presents the relation between the metric and first-order formalisms, while the second summarizes the framework used to compute black hole entropy.

\subsection{Relation Between Metric and First-Order Formalisms}

In GR, \( g_{ab} \) is the dynamical field describing gravity, and the connection is uniquely determined by it. This is known as the Levi-Civita connection. It is, however, possible to consider a connection that is dynamically independent. The properties of such a connection are closely tied to those of a derivative operator \( \nabla_a \), as the connection specifies its action on tensor fields~\cite{WaldB}. In general, a connection can be decomposed into a metric-compatible (Levi-Civita) part, a nonmetricity contribution, and a torsion contribution.

The nonmetricity tensor is defined as \( Q_{abc} = \nabla_a g_{bc} \) and characterizes the failure of the connection to preserve lengths under parallel transport. Since nonmetricity may lead to pathological behavior~\cite{Pala2023}, it is set to zero in what follows (alternatively, it may be treated as a matter field). The torsion tensor, \( T^{c}{}_{ab} \), quantifies the nonclosure of infinitesimal parallelograms~\cite{Naka} and is defined through
\begin{equation}
(\nabla_{a} \nabla_{b} - \nabla_{b} \nabla_{a}) f = - T^{c}{}_{ab} \nabla_{c} f,
\label{torsion}
\end{equation}
for a generic function $f$. Clearly, \( T^{c}{}_{ab} = -T^{c}{}_{ba} \). It is convenient to introduce the contorsion tensor, 
\begin{equation}
K^{c}{}_{ab} = \frac{1}{2} \left( T^{c}{}_{ab} - T_{a}{}^{c}{}_{b} - T_{b}{}^{c}{}_{a} \right),
\label{Contor}
\end{equation}
which links the Levi-Civita derivative with \( \nabla_a \)~\cite{NesterTor}.

The Riemann tensor associated with \( \nabla_a \) can be expressed as
\begin{eqnarray}
{R_{abc}}{}^{d} &=& {\mathring{R}_{abc}}{}^{d} - \mathring{\nabla}_a K_{bc}{}^{d} + \mathring{\nabla}_b K_{ac}{}^{d} \nonumber \\
&&+ K^{e}{}_{ac} K_{be}{}^{d} - K^{e}{}_{bc} K_{ae}{}^{d},
\label{Riemanns}
\end{eqnarray}
where a ring denotes Levi-Civita quantities. One has \( {R_{abc}}{}^{d} = -{R_{bac}}{}^{d} \) and \( R_{abcd} = -R_{abdc} \), while in general \( {R_{[abc]}}{}^{d} \neq 0 \). Here, square brackets denote antisymmetrization (with a factor of one over the factorial of the number of antisymmetrized indices).

When using differential forms, the fundamental variables are the vielbein 1-forms \( e^{\mu} \) (tetrads in four dimensions) and the spin connection 1-form \( \omega^{\mu\nu} = -\omega^{\nu\mu} \), which, for the sake of generality, are treated as independent fields. These satisfy~\cite{Ruso}
\begin{equation}
g_{ab} = \eta_{\mu \nu} \, e^{\mu}_{a} \, e^{\nu}_{b},
\label{Metrictetrad}
\end{equation}
and
\begin{equation}
\omega^{\mu}{}_{\nu} = \Gamma^{\mu}{}_{\alpha \nu} \, e^{\alpha},
\label{connectiontrans}
\end{equation}
where \( \Gamma^{\mu}{}_{\alpha \nu} \) are the components, in the vielbein basis, of the connection linking a partial derivative and \( \nabla_a \).

The curvature 2-form \( R^{\mu\nu} \) is obtained by contracting the last two indices of the Riemann tensor with vielbeins. In terms of the spin connection, it is given by
\begin{equation}
R^{\mu\nu} = \de \omega^{\mu\nu} + \omega^{\mu}{}_{\alpha} \wedge \omega^{\alpha\nu},
\label{2curvaform}
\end{equation}
where \( \de \) is the exterior derivative and \( \wedge \) denotes the wedge product~\cite{Naka}. The spin connection admits the decomposition
\begin{equation}
\omega^{\mu}{}_{\nu} = \mathring{\omega}^{\mu}{}_{\nu} + K^{\mu}{}_{\nu},
\label{decomposicion}
\end{equation}
where \( \mathring{\omega}^{\mu}{}_{\nu} \) is associated with the Levi-Civita connection \( \mathring{\Gamma}^{\mu}{}_{\alpha \nu} \),
\begin{equation}
\mathring{\omega}^{\mu}{}_{\nu} = \mathring{\Gamma}^{\mu}{}_{\alpha \nu} \, e^{\alpha},
\label{connectiontransRing}
\end{equation}
and \( K^{\mu}{}_{\nu} = K^{\mu}{}_{\rho\nu} e^{\rho} \), with $K^{\mu}{}_{\rho\nu}$ the components of the contorsion in the vielbein basis.

In \( n \) spacetime dimensions, the Hodge dual \( \star \) maps a \( p \)-form into an \( (n{-}p) \)-form. It acts as
\begin{equation}\label{Hodge}
\star(e^{\mu_1} \wedge \cdots \wedge e^{\mu_p}) = \frac{1}{(n{-}p)!} \epsilon^{\mu_1 \dots \mu_p}{}_{\mu_{p+1} \dots \mu_n} \, e^{\mu_{p+1}} \wedge \cdots \wedge e^{\mu_n},
\end{equation}
where \( \epsilon_{\mu_1 \dots \mu_n} \) is the Levi-Civita symbol.

The covariant exterior derivative \( \text{D} \) reduces to \( \de \) when acting on group scalars. However, when group indices are present, it includes a spin-connection term for each such index (with a sign depending on the position of such an index). For instance,
\begin{equation}
\text{D} e^\mu = \de e^\mu + \omega^{\mu}{}_{\nu} \wedge e^\nu,
\end{equation}
which defines the torsion 2-form \( T^\mu \). In the vielbein basis,
\begin{equation}
T^\mu = \frac{1}{2} T^\mu{}_{\alpha\beta} \, e^\alpha \wedge e^\beta,
\end{equation}
where \( T^\mu{}_{\alpha\beta} \) are the torsion components. Thus,
\begin{equation}\label{omegaring}
\de e^\mu + \mathring{\omega}^{\mu}{}_{\nu} \wedge e^\nu = 0.
\end{equation}

Finally, it is possible to show that the Bianchi identity, in terms of the covariant exterior derivative, is
\begin{equation}\label{Bianchi}
\text{D} R^{\mu\nu} =0.
\end{equation}
One can also show that
\begin{equation}\label{BianchiT}
\text{D} T^{\mu} =R^{\mu\nu}\wedge e_\nu.
\end{equation}

\subsection{Geometric Derivation of Black Hole Entropy}\label{WaldDerivation}

For a diffeomorphism-invariant theory, the first law of black hole thermodynamics is related to Noether’s second theorem~\cite{wald}. In this setup, the Lagrangian $\mathcal{L}$ is an $n$-form depending on the dynamical fields, collectively denoted by $\phi$, and their (symmetrized) covariant derivatives~\cite{Iyer}. Hence, its variation takes the general form
\begin{equation}
 \delta \mathcal{L} = \delta \phi \wedge \mathcal{E}_\phi + \de \theta(\delta \phi),
 \label{WaldVarofL}
\end{equation}
where $\mathcal{E}_\phi=0$ are the equations of motion (one for each independent field), and the $(n-1)$-form  $\theta(\delta \phi)$ is the boundary term (which depends on the fields but also on their variations, as indicated explicitly by the notation).

A diffeomorphism is generated by the Lie derivative along a vector field $\xi$, $\mathcal{L}_\xi = \de\, \mathrm{i}_\xi + \mathrm{i}_\xi \de$, where $\mathrm{i}_\xi$ denotes contraction; this expression for the Lie derivative when acting on differential forms is known as Cartan's magic formula. Equation~\eqref{WaldVarofL} can then be written as
\begin{equation}
\de \mathrm{i}_{\xi} \mathcal{L} = \mathcal{L}_\xi \phi \wedge \mathcal{E}_\phi + \de \theta(\mathcal{L}_\xi \phi).
 \label{CartanMagic}
\end{equation}
Then, on shell, there exists an $(n-1)$-form,
\begin{equation}
 J(\xi) = \theta(\mathcal{L}_\xi \phi) - \mathrm{i}_{\xi} \mathcal{L},
 \label{Jo}
\end{equation}
such that $\de J=0$. Moreover, $J$ has an associated Noether charge $Q$ satisfying $J = \de Q$, which exists by the Poincaré lemma (at least locally).

The Noether current is directly related to the presymplectic current, defined as~\cite{Simone,Prabhu_2017}
\begin{equation}
 \Omega(\delta_{1},\delta_{2}) 
 = \int_\Sigma \delta_{1} \theta(\delta_{2}) 
 - \delta_{2} \theta(\delta_{1}) 
 - \theta([\delta_{1},\delta_{2}]),
 \label{Symplectic}
\end{equation}
where $\delta_1$ and $\delta_2$ are variations of the dynamical fields, and the integral is taken over a Cauchy hypersurface $\Sigma$. If $\delta_{1}$ is a variation within the space of solutions to the field equations and $\delta_{2} = \mathcal{L}_{\xi}$, the variations commute, yielding
\begin{equation}
\Omega(\delta, \mathcal{L}_{\xi}) 
= \int_\Sigma \de \left[\delta Q(\xi) 
-  \mathrm{i}_{\xi}\theta(\delta)\right] = \int_{\partial\Sigma} \delta Q(\xi) 
-  \mathrm{i}_{\xi}\theta(\delta),
\label{firstlawpreSym-bdry3}
\end{equation}
where Stokes' theorem is used and $\partial\Sigma$ denotes the boundary of $\Sigma$.

Now, for a Killing vector field $\xi$,
\begin{equation}
  0=\mathcal{L}_\xi g_{ab} =-2\mathring{\nabla}_{(a}\xi_{b)}.   
\end{equation}
However, in a theory where all dynamical fields couple with each other, all dynamical fields must be invariant under this flow. Therefore,
\begin{equation}
\Omega(\delta, \mathcal{L}_\xi) = 0.
\end{equation}

\begin{figure}[h]
\centering
\includegraphics[width=0.45\textwidth]{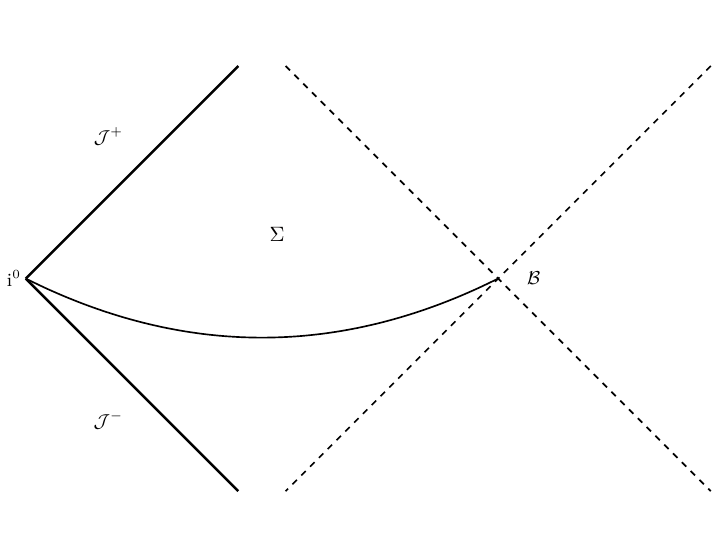}
\label{fig}
\caption{Penrose diagram of a stationary black hole with a bifurcate Killing horizon. The spacetime is asymptotically flat, and standard notation is used: $\text{i}^0$ denotes spatial infinity, while $\mathcal{J}^\pm$ denote future and past null infinity. The exterior region is globally hyperbolic and admits a foliation by Cauchy surfaces, one of the is denoted by $\Sigma$.}
\end{figure}

Notice that spacetime is assumed to be asymptotically flat with a globally hyperbolic exterior region~\cite{WaldB}, ensuring the existence of a Cauchy surface $\Sigma$ for this region. It is further assumed that a bifurcate Killing horizon generated by $\xi$ is present. A Penrose diagram of such a spacetime is shown in Fig.~\ref{fig}. In this case $\Sigma$ has two boundaries: spatial infinity $\text{i}^0$ and the bifurcation surface $\mathcal{B}$~\cite{WaldB}. Equation~\eqref{firstlawpreSym-bdry} then becomes
\begin{equation}
\int_{\mathcal{B}} \delta Q(\xi) = \int_{\text{i}^0} \delta Q(\xi) 
-  \mathrm{i}_{\xi}\theta(\delta),
\label{firstlawpreSym-bdry}
\end{equation}
where the fact that $\xi=0$ on $\mathcal{B}$ is used. This last equation is the first law of black hole thermodynamics.

On the Killing horizon, and given that $\xi$ is orthogonal to it, 
\begin{equation}
\xi^b \mathring{\nabla}_b \xi_a =-\xi^b \mathring{\nabla}_a \xi_b = \chi \xi_a,
\end{equation}
where $\chi$ is the surface gravity, which is constant over $\mathcal{B}$. It can then be argued that $\mathring{\nabla}_a \xi_b$ has no components tangent to $\mathcal{B}$ and is therefore proportional to the binormal 2-form $n = n_{\mu \nu} e^\mu \wedge e^\nu/2$, whose Hodge dual defines the volume element of $\mathcal{B}$. The integral over $\mathcal{B}$ in Eq.~\eqref{firstlawpreSym-bdry} thus contains a term proportional to $\chi$, identified with the black hole temperature, and a term proportional to the horizon volume, identified with the entropy. Moreover, the integral over $\text{i}^0$ yields the Hamiltonian charges, that include the internal energy and other thermodynamic potentials.

\section{Gauge-Independent Prescription for the First Law of Black Hole Thermodynamics}\label{BHEFOF}

In this section, the formalism for deriving the first law of black hole thermodynamics is reformulated in the language of differential forms. This requires several modifications. In particular, the presymplectic current must be corrected to account for the fact that now gravity is described as a gauge theory. In fact, together with diffeomorphisms, gauge symmetries form a closed algebra~\cite{review,BonCris,Monte}, whose interpretation neatly emerges at the level of the associated principal fiber bundle~\cite{Algebrasym}.

\subsection{Gauge charges}

The key difference is the presence of gauge (GT) charges. These arise from
\begin{equation} 
0 = \delta_{\text{GT}}(\lambda) \mathcal{L} = \de \theta(\delta_{\text{GT}}(\lambda)),
\end{equation}
where the first equality follows from the fact that the Lagrangian is a group scalar, and the second from the general on-shell variation of the Lagrangian. By the same arguments as above, there exists a charge $Q_{\text{GT}}$ such that
\begin{equation} \label{betaorig}
\theta(\delta_{\text{GT}}(\lambda)) = \de Q_{\text{GT}}(\lambda),
\end{equation}
where $\lambda$ is the parameter of the gauge transformation (i.e., it can be ``field-dependent'' in the terminology of Ref.~\onlinecite{Lee}).

Importantly, in the first-order formalism, the definition of a Killing vector field $\xi$ requires $\mathcal{L}_{\xi} e^{\mu}$ to be a gauge transformation of $e^\mu$, whose corresponding gauge parameter is derived in Appendix~\ref{GaugeParam}. Thus, requiring $\Omega(\delta, \delta_{\text{Diff}}(\xi)) = 0$ for a Killing vector field, as in the metric formalism, is overly restrictive. Furthermore, for this diffeomorphism to be a symmetry, it is necessary that the Lie derivatives of the other fields along $\xi$ vanish. Under this assumption, for any of the fields under consideration, the actions of $\delta_{\text{Diff}}(\xi)$ and $\delta_{\text{GT}}(\lambda_\xi)$, with $\lambda_\xi$ satisfying Eq.~\eqref{GaugeTKilling}, are equivalent. Accordingly, the following identity holds at the level of transformations:
\begin{equation}\label{diffKilling}
\delta_{\text{Diff}}(\xi) = \delta_{\text{GT}}(\lambda_\xi).
\end{equation}
Then, it is then natural to assume (cf. Ref.~\onlinecite{Simone})
\begin{equation}\label{nonzeroxymplectic}
\Omega(\delta, \delta_{\text{Diff}}(\xi)) = \Omega(\delta, \delta_{\text{GT}}(\lambda_{\xi})).
\end{equation}

The right-hand side of Eq.~\eqref{nonzeroxymplectic} satisfies
\begin{equation} \label{gtSymplectic}
\Omega(\delta,\delta_{\text{GT}}(\lambda_{\xi})) = \int_\Sigma \delta \theta (\delta_{\text{GT}} (\lambda_{\xi})) - \theta( \delta_{\text{GT}} (\delta \lambda_\xi) ),
\end{equation}
where it is used that $\theta(\delta)$ is a group scalar, i.e., $\delta_{\text{GT}}(\lambda_{\xi})\, \theta(\delta) = 0$, and
\begin{equation}\label{conmutador}
 [\delta,\delta_{\text{GT}}(\lambda)] = \delta_{\text{GT}} (\delta \lambda),
\end{equation}
as $\lambda^{\mu\nu}$ may be field dependent. Moreover, Eq.~\eqref{gtSymplectic} can be written as
\begin{equation} \label{gtSymplecticbeta}
\Omega(\delta,\delta_{\text{GT}}(\lambda_{\xi})) = \int_\Sigma \de \left[\delta Q_\text{GT} (\lambda_{\xi}) - Q_\text{GT}( \delta \lambda_\xi )\right],
\end{equation}
where the fact that $\de$ and $\delta$ commute is applied. Finally, inserting these results into Eq.~\eqref{nonzeroxymplectic} and using Stokes' theorem gives
\begin{equation}\label{SympCurrentJStokes}
 0 = \oint_{\partial\Sigma} \delta Q(\xi) - \,\text{i}_\xi \theta(\delta) - \delta Q_{\text{GT}}(\lambda_{\xi})
 + Q_{\text{GT}}(\delta \lambda_\xi).
\end{equation}

\subsection{Gauge independence}

A natural question is whether the first law of black hole thermodynamics can also be derived using a more general symmetry transformation that combines a diffeomorphism with an arbitrary gauge transformation. Specifically, consider a transformation of the form
\begin{equation}
\delta (\Lambda) = \delta_{\text{Diff}}(\xi) - \delta_{\text{GT}}(\Lambda),
\label{GT+Diff}
\end{equation}
where $\Lambda$ is a possibly field-dependent gauge parameter, which may also depend on $\xi$. When $\xi$ is a Killing vector field, Eq.~\eqref{diffKilling} applies, and
\begin{equation}
 \delta (\Lambda) 
 = \delta_{\text{GT}}(\lambda_\xi)-\delta_{\text{GT}}(\Lambda)
 = \delta_{\text{GT}}(\lambda_\xi-\Lambda),
 \label{GT+Diff1}
\end{equation}
where the last equality follows from the properties of the Lie group. Then, as before, the first law of black hole thermodynamics should be derived from
\begin{equation}
 \Omega(\delta, \delta (\Lambda))
 = \Omega(\delta, \delta_{\text{GT}}(\lambda_\xi-\Lambda)).
 \label{GT+Diff2}
\end{equation}

To demonstrate that this transformation yields the same first law, a property known as gauge independence, let $\delta_1$ and $\delta_2$ denote arbitrary diffeomorphisms or gauge transformations. Then, for any real coefficient $\alpha$, 
\begin{equation}\label{linearity}
\Omega(\delta, \delta_1 + \alpha \delta_2) = \Omega(\delta, \delta_1) + \alpha \, \Omega(\delta, \delta_2),
\end{equation}
and the left-hand side of Eq.~\eqref{GT+Diff2} satisfies
\begin{equation}\label{31}
 \Omega(\delta, \delta (\Lambda) )
 = \Omega(\delta, \delta_{\text{Diff}}(\xi)) 
 - \Omega(\delta, \delta_{\text{GT}}(\Lambda)).
\end{equation}
Similarly, the right-hand side becomes
\begin{equation}
 \Omega(\delta, \delta_{\text{GT}}(\lambda_\xi-\Lambda))
 = \Omega(\delta, \delta_{\text{GT}}(\lambda_\xi)) 
 - \Omega(\delta, \delta_{\text{GT}}(\Lambda)).
\end{equation}
A comparison of the last two expressions shows that the terms $\Omega(\delta, \delta_{\text{GT}}(\Lambda))$ cancel, yielding Eq.~\eqref{nonzeroxymplectic} and thereby reproducing the ``original'' first law of black hole thermodynamics. Hence, when formulated with Eq.~\eqref{GT+Diff2}, the first law of black hole thermodynamics is gauge independent.

In Ref.~\onlinecite{Simone}, Einstein--Cartan theory is analyzed, and it is argued that, for the presymplectic current to be gauge independent, $\theta$ must be modified by the addition of an \emph{ad hoc} exact form. Here, it is shown that this additional contribution arises naturally. In addition, the framework presented here applies to arbitrary theories. The key observation is that, since $\theta$ is linear, and since both the gauge transformations and the exterior derivative are linear as well, $Q_{\text{GT}}(\lambda)$ must also be linear.

Note that one can always write
\begin{equation}
 \lambda^{\mu \nu} = e^{[\mu}_a \delta_{\text{GT}} (\lambda) e^{\nu] a}.
 \label{Gtgamma}
\end{equation}
This suggests the following definition:
\begin{equation}
 \gamma (\delta) = - Q_\text{GT}\bigl(e^{[\mu}_a \delta e^{\nu] a}\bigr) ,
 \label{Gamma}
\end{equation}
where $\delta$ is an arbitrary variation. This object has the property that in the particular case $\delta = \delta_{\text{GT}}(\lambda)$, one has
\begin{equation}
 \gamma (\delta_\text{GT}(\lambda)) = - Q_\text{GT}(\lambda).
 \label{GammaParticular}
\end{equation}
Linearity implies that there exists an antisymmetric co-dimension 2 tensor field $H_{\mu\nu}$, to be determined below, such that
\begin{equation}\label{beta1}
 Q_{\text{GT}}(\lambda) = \lambda^{\mu\nu} H_{\mu\nu}.
\end{equation}
Furthermore, $H_{\mu\nu}$ depends on the dynamical fields and transforms covariantly under gauge transformations. Hence,
\begin{equation}
 \gamma(\delta) = - H_{\mu\nu} \, e^\mu{}_a \, \delta e^{\nu a}.
\end{equation}

Now, one can consider an arbitrary variation of $\lambda_\xi^{\mu\nu}$, defined in Eq.~\eqref{GaugeTKilling}, and use it as the argument of $Q_{\text{GT}}$. This yields
\begin{eqnarray}
 Q_{\text{GT}}(\delta\lambda_\xi)
 &=& (\lambda_\xi)^\mu{}_{\rho} (e^\rho_a \delta e^{\nu a}) H_{\mu \nu}
 + (\lambda_\xi)^\nu{}_{\rho} (e^\mu_a \delta e^{\rho a}) H_{\mu \nu} \nonumber\\
  && + \mathcal{L}_\xi (e^\mu{}_a \delta e^{\nu a}) H_{\mu \nu}.
 \label{betadeltalambda}
\end{eqnarray}
The first two terms correspond to a gauge transformation of $\gamma$ that vanish by virtue of the fact that $\gamma$ is a group scalar. In addition, since $H_{\mu\nu}$ transforms covariantly, its Lie derivative along the Killing field $\xi$ corresponds to a gauge transformation. Consequently, Eq.~\eqref{betadeltalambda} can be rewritten as
\begin{equation}
 Q_{\text{GT}}(\delta\lambda_\xi) = -\mathcal{L}_\xi \gamma(\delta) = -\text{i}_\xi \de \gamma(\delta) -\de \text{i}_\xi \gamma(\delta),
 \label{betadeltalambda2}
\end{equation}
where Cartan's magic formula is utilized. Substituting this result into Eq.~\eqref{gtSymplecticbeta} demonstrates that the standard symplectic current picks up boundary contributions due to the Lorentz gauge freedom. To absorb these terms and restore an expression of the first law similar to that of the metric formalism, it is convenient to define the ``corrected'' boundary term and Noether charge:
\begin{eqnarray}
 \hat{\theta}(\delta) = \theta(\delta) + \de \gamma(\delta), \quad
 \hat{Q}(\xi) = Q(\xi) - Q_{\text{GT}}(\lambda_\xi) .
 \label{ChargeHQ}
\end{eqnarray}
Consider now the general transformation given in Eq.~\eqref{GT+Diff}. The associated Noether current is
\begin{equation}
J(\Lambda,\xi) = \theta(\delta (\Lambda)) - \text{i}_\xi \mathcal{L} = J(\xi) - \theta_{\text{GT}}(\Lambda).
\label{NoetherCurrentATR}
\end{equation}
The corresponding Noether charge then takes the form
\begin{equation}
Q (\Lambda, \xi) = Q (\xi) - Q_{\text{GT}} (\Lambda),
\label{ChargeATR}
\end{equation}
where Eq.~\eqref{betaorig} is applied. It follows that
\begin{equation}
\hat{Q} (\xi) = Q (\Lambda, \xi) - Q_{\text{GT}}(\lambda_\xi - \Lambda) = Q(\xi) - Q_{\text{GT}}(\lambda_\xi).
\label{QhatT}
\end{equation}
Note that all terms depending on $\Lambda$ cancel out, showing that $\hat{Q}(\xi)$ is independent of the choice of gauge. Likewise, the gauge independence of the boundary term $\hat{\theta}(\delta)$ follows directly from Eq.~\eqref{GammaParticular}.

These modified quantities provide the necessary ingredients to construct a fully gauge-independent formulation of the presymplectic current, as shown below.

\subsection{Gauge independent presymplectic current}
It is noteworthy that $\hat{\theta}(\delta)$ allows the construction of a ``gauge-independent presymplectic current'' by replacing $\theta(\delta)$ with $\hat{\theta}(\delta)$ in Eq.~\eqref{Symplectic}. This is the generalization of the presymplectic current proposed in Ref.~\onlinecite{Simone} for Einstein--Cartan theory. Accordingly,
\begin{eqnarray}
 \hat{\Omega} (\delta_1,\delta_2) &=& \int_\Sigma \delta_1 \hat{\theta} (\delta_2) - \delta_2 \hat{\theta} (\delta_1) - \hat{\theta}([\delta_1,\delta_2]) \nonumber \\
 &=& \Omega(\delta_1,\delta_2) + \int_\Sigma \de  \Gamma(\delta_1,\delta_2),
 \label{PreSGT}
\end{eqnarray}
where
\begin{equation}
\Gamma(\delta_1,\delta_2)= \delta_1 \gamma(\delta_2) 
  - \delta_2 \gamma(\delta_1) - \gamma([\delta_1,\delta_2]).
 \end{equation}

The gauge-independent presymplectic current has two interesting properties. First, it vanishes for a gauge transformation. Namely,
\begin{equation}\label{OmegaHatGauge}
 \hat{\Omega} (\delta, \delta_\text{GT}(\lambda))=0. 
\end{equation}
The proof follows directly from Eqs.~\eqref{conmutador},~\eqref{gtSymplecticbeta} and \eqref{GammaParticular}, together with the fact that $\gamma$ is a group scalar. The second property is that it is, in fact, gauge independent, which is expressed as
\begin{equation}
 \hat{\Omega} (\delta, \delta(\Lambda)) = \hat{\Omega}(\delta, \mathcal{L}_\xi).
 \label{IndGT1st}
\end{equation}
This property follows directly from the linearity of $\hat{\Omega} (\delta_1, \delta_2)$ and Eq.~\eqref{OmegaHatGauge}.

Most importantly, when $\xi$ is a Killing vector field, Eqs.~\eqref{OmegaHatGauge} and \eqref{IndGT1st} imply that the first law of black hole thermodynamics follows from
\begin{equation}\label{FirstLawHat}
 0=\hat{\Omega} (\delta,\mathcal{L}_\xi) .
\end{equation}
This is because, on the left-hand side, any gauge transformation can be considered, yet it vanishes; and on the right-hand side, any transformation is allowed, but the result is equivalent. Furthermore, Eq.~\eqref{FirstLawHat} closely parallels the condition used in the metric formalism: $0=\Omega(\delta,\mathcal{L}_\xi)$. The proof that Eq.~\eqref{FirstLawHat} reduces to the first law of black hole thermodynamics relies on the fact that $\delta$ and $\mathcal{L}_\xi$ commute, and thus, 
\begin{equation}
 \hat{\Omega} (\delta,\mathcal{L}_\xi) 
= \Omega(\delta,\mathcal{L}_\xi) + \int_\Sigma \de \left[ \delta \gamma(\mathcal{L}_\xi)- \mathcal{L}_\xi \gamma(\delta) \right] . \label{PreSGT2}
\end{equation}
Then, by using Eqs.~\eqref{firstlawpreSym-bdry3} and \eqref{QhatT}, one can bring Eq.~\eqref{PreSGT2} to a form that is analogous to Eq.~\eqref{firstlawpreSym-bdry3}.

The definition of $\hat{\Omega}(\delta_1,\delta_2)$ constitutes one of the main proposals of this work. Notably, although the Killing condition in first-order formalisms no longer requires strict invariance of all fields under the flow, but only invariance up to a gauge transformation (as is the case for the vielbein), our prescription ensures that $\hat{\Omega} (\delta, \mathcal{L}_\xi) = 0$ holds independently of the gauge transformation involved. This leads directly to the gauge-invariant first law of black hole thermodynamics:
\begin{equation}
 \int_{\mathcal{B}} \delta \hat{Q}(\xi)  = \int_{\text{i}^0} \delta \hat{Q}(\xi) - \text{i}_\xi \hat{\theta}(\delta) .
 \label{1stlawcorrected}
\end{equation}

From a practical standpoint, Eq.~\eqref{1stlawcorrected} provides a direct and fully gauge-independent prescription to compute the first law of black hole thermodynamics, working only with objects defined in spacetime (cf. Ref.~\onlinecite{Prabhu_2017}). Evaluating Eq. \eqref{1stlawcorrected} in a general theory constitutes another goal of this work, which is studied in the following sections. 

\section{Arbitrary Theories of Gravity} \label{SecGTGl}

\subsection{Dynamics}

The focus of this section is on how to write the action of a general gravity theory using differential forms. Recall that the action must be invariant under gauge transformations and diffeomorphisms (perhaps up to a boundary term). This invariance can be achieved by not allowing for nondynamical tensors \cite{Iyer,BonCris}. Thus, the fundamental tensors under consideration are the vielbeins and the spin connections. In addition, the theory has to be covariant. Hence, the spin connection $\omega^{\mu \nu}$ can only appear through curvature or torsion, which is the covariant exterior derivative of the vielbein. Moreover, the Bianchi identities, Eqs.~\eqref{Bianchi} and \eqref{BianchiT}, and the fact that, for any algebra-valued $p$-form $U^{\mu_1 \dots \mu_p}$,
\begin{equation}
    \D \D U^{\mu_1 \dots \mu_p} = R^{\mu_1}_{\ \ \ \alpha} \wedge U^{\alpha \dots \mu_p} + \dots +R^{\mu_p}_{\ \ \ \alpha} \wedge U^{\mu_1 \dots \alpha},
    \label{doubleD}
\end{equation}
(and a similar expression holds when the group indices are in the lower position), allow one to omit exterior covariant derivatives of curvature and torsion, as well as higher covariant exterior derivatives, from the functional dependence of the Lagrangian. Hence, to include higher order derivatives \cite{Zanelli:2005sa} it is necessary to introduce the Hodge star operator (bearing in mind that  $\D \star e^{\mu}$ is encoded through torsion). In light of these considerations, a general gravity theory is described by 
\begin{widetext}
\begin{equation}
 S [e,\omega] = \int \ \mathcal{L}_G[e, R, \star R, T, \star T,   \D \star R, \star \D \star R, \dots,\D^{N} \star R, \star \D^{N} \star R, \ \D \star T, \star \D \star T, \dots, \D^{M} \star T, \star \D^{M} \star T],
 \label{TheL}
\end{equation}
\end{widetext}
where $\D^{l} = \D \star \D \star \dots \star \D $, with $l$ factors of $\D$ and with the convention that $\D^{1} = \D$, and $N$ and $M$ denote the maximum number of derivatives acting on curvature and torsion, respectively. Moreover, given that boundary contributions do not affect the first law of black hole thermodynamics (see Appendix~\ref{InvLag}), it is assumed that the Lagrangian contains no exact terms. Notice that the subindex $G$ stands for gravity and should not be confused with GT.

\subsection{Equations of motion and boundary term}

An arbitrary variation of Eq.~\eqref{TheL} can be expressed as
\begin{widetext}
\begin{eqnarray}
 \delta S &=& \int \ \delta e^{\mu} \wedge \frac{\partial \mathcal{L}_G}{\partial e^{\mu}} + \delta R^{\mu \nu} \wedge \frac{\partial \mathcal{L}_G}{\partial R^{\mu \nu}} + \delta T^{\mu} \wedge \frac{\partial \mathcal{L}_G}{\partial T^{\mu}}  +\delta \star R^{\mu \nu} \wedge \frac{\partial \mathcal{L}_G}{\partial \star R^{\mu \nu}} + \delta \star T^{\mu} \wedge \frac{\partial \mathcal{L}_G}{\partial \star T^{\mu}} \nonumber \\
    && \ + \sum_{j=1}^{N} \delta \star \D^{j}  \star R^{\mu \nu} \wedge  \frac{\partial \mathcal{L}_G}{\partial \star \D^{j}  \star R^{\mu \nu}} +   \sum_{j=1}^{N} \delta  \D^{j}  \star R^{\mu \nu} \wedge  \frac{\partial \mathcal{L}_G}{\partial \D^{j}  \star R^{\mu \nu}} \nonumber \\
    &&  \ + \sum_{j=1}^{M} \delta \star \D^{j}  \star T^{\mu} \wedge  \frac{\partial \mathcal{L}_G}{\partial \star \D^{j}  \star T^{\mu }} +   \sum_{j=1}^{M} \delta  \D^{j}  \star T^{\mu} \wedge  \frac{\partial \mathcal{L}_G}{\partial \D^{j}  \star T^{\mu}}.
 \label{GVequation}
\end{eqnarray}
\end{widetext}
where following identities are repeatedly utilized:
\begin{eqnarray}
 \delta \star U^{\mu_{1} \dots \mu_{n}} &=& \star \delta U^{\mu_{1} \dots \mu_{n}} - \star[\delta e^{\alpha} \wedge\text{i}_{\alpha} U^{\mu_{1} \dots \mu_{n}}] \nonumber\\ 
 && + \delta e^{\alpha} \wedge \text{i}_{\alpha} \star U^{\mu_{1} \dots \mu_{n}},
 \label{Idstar}
\end{eqnarray}
\begin{eqnarray}
 \delta \D U^{\mu_1 \dots \mu_n} &=& \D \delta U^{\mu_1 \dots \mu_n} + \delta \omega^{\mu_1}_{\ \alpha} \wedge U^{\alpha \dots \mu_n} + \dots \nonumber \\
 &&+ \delta \omega^{\mu_n}_{\ \alpha} \wedge U^{\mu_1 \dots \alpha},
 \label{VarDU}
\end{eqnarray}
with $U^{\mu_{1} \dots \mu_{n}}$ denoting a generic $p$-form depending on the dynamical fields. 

Let
\begin{equation}
    I^{(N)}_{\mu \nu} = \frac{\partial \mathcal{L}_G}{\partial \star \D^{N} \star R^{\mu \nu}}, 
\end{equation}
for $N \geq 1$, and consider the variation of the highest-derivative term, which satisfies
\begin{eqnarray}
 \delta \left(\star \D^{N} \star R^{\mu \nu} \right) \wedge I^{(N)}_{\mu \nu} &=& \star \left( \delta\D^{N} \star R^{\mu \nu} \right)  \wedge I^{(N)}_{\mu \nu}  \nonumber \\
 &&-\star \left( \delta e^{\alpha} \wedge \text{i}_\alpha\D^{N} \star R^{\mu \nu} \right) \wedge  I^{(N)}_{\mu \nu} \nonumber \\
 && + \delta e^{\alpha} \wedge \text{i}_\alpha \star \D^{N} \star R^{\mu \nu}  \wedge I^{(N)}_{\mu \nu}, \nonumber \\
 \label{VarofDN2}
\end{eqnarray}
where $\text{i}_\mu $ represents the contraction with the vector field dual to $e^\mu$, and Eqs.~\eqref{Idstar}-\eqref{VarDU} are used. The last two terms in Eq.~\eqref{VarofDN2} contribute to the equations of motion for $e^{\mu}$, while the first term can be rewritten, using the properties of the Hodge operator~\cite{Naka}, which allow one to write
\begin{equation}
 \star \left( \delta\D^{N} \star R^{\mu \nu} \right) \wedge I_{\mu \nu}^{(N)} = (-1)^{N(n+1)} \delta\D^{N} \star R^{\mu \nu}  \wedge \star I_{\mu \nu}^{(N)}.
    \label{I25}
\end{equation}
Then, the terms with $\delta\D^{N} \star R^{\mu \nu}$ in $\delta S$ can be grouped as $\delta\D^{N} \star R^{\mu \nu} \wedge H_{\mu \nu}^{(N)}$, where
\begin{equation}
    H_{\mu \nu}^{(N)}= \frac{\partial \mathcal{L}_G}{\partial \D^{N} \star R^{\mu \nu}} +(-1)^{N(n+1)}  \star I_{\mu \nu}^{(N)}. 
    \label{H1} 
\end{equation}

On the other hand, the contribution of $\D^{N} \star R^{\mu \nu}$ can be expressed as
\begin{widetext}
\begin{eqnarray}
 \delta \left( \D^{N} \star R^{\mu \nu} \right) \wedge H_{\mu\nu}^{(N)} &=&
  \de \left[ \delta \left( \star  \D^{N-1} \star R^{\mu \nu} \right) \wedge H_{\mu\nu}^{(N)} \right] 
 + (-1)^{N(n+1)}   \delta\star \D^{N-1} \star R^{\mu \nu}  \wedge \D H_{\mu \nu}^{(N)} \nonumber \\
 && + 2 \delta \omega^{[\mu}_{\ \alpha} \wedge \star \D^{N-1} \star R^{|\alpha| \nu]} \wedge H_{\mu\nu}^{(N)}.
 \label{VarofDN}
\end{eqnarray}
\end{widetext}
From these expressions, one can then read off the contributions to the equation of motion and the corresponding boundary contribution.

By recursion, the variations $\delta\star \D^{N-1} \star R^{\mu \nu}$ and $\delta \D^{N-1} \star R^{\mu \nu}$ can be expressed as
\begin{eqnarray}
    I_{\mu \nu}^{(N-1)} &=& \frac{\partial \mathcal{L}_G}{\partial \star \D^{N-1} \star R^{\mu \nu}}+ (-1)^{N(n+1)}    \D H_{\mu \nu}^{(N)},  \label{curvaturen-1} \\
    H_{\mu \nu}^{(N-1)} &=&  \frac{\partial \mathcal{L}_G}{\partial \D^{N-1} \star R^{\mu \nu}} + (-1)^{(N-1)(n+1)} \star I_{\mu \nu}^{(N-1)}. \nonumber \\  
    \label{HGTNHO0}
\end{eqnarray}
In this way, the variation of the highest-order term, $\D^{N} \star R^{\mu \nu}$, induces contributions to the variation of the $(N-1)$-derivative term, whose structure in turn propagates to lower orders. The lowest order is
\begin{eqnarray}
    I_{\mu \nu}^{(0)} &=& \frac{\partial \mathcal{L}_G}{\partial  \star R^{\mu \nu}}- (-1)^{N}    \D H_{\mu \nu}^{(1)},  \\
    H_{\mu \nu}^{(0)} &=&  \frac{\partial \mathcal{L}_G}{\partial R^{\mu \nu}} + \star \frac{\partial \mathcal{L}_G}{\partial \star R^{\mu \nu}} - (-1)^{N} \star \D H_{\mu \nu}^{(1)}.  
    \label{Hzero}
\end{eqnarray}

A similar scheme can be applied to torsion and its derivatives. For the highest derivative order, $M$, one has
\begin{eqnarray}
    I_{\mu }^{(M)} &=& \frac{\partial \mathcal{L}_G}{\partial \star \D^{M} \star T^{\mu}}, \label{TorsioI} \\
    E_{\mu }^{(M)} &=&  \frac{\partial \mathcal{L}_G}{\partial \D^{M} \star T^{\mu }} + (-1)^{M(n+1)} \star I_{\mu}^{(n)}.   
    \label{TorsionHE}
\end{eqnarray}
Thus, 
\begin{eqnarray}
    I_{\mu }^{(M-1)} &=& \frac{\partial \mathcal{L}_G}{\partial \star \D^{M-1} \star T^{\mu}}+ (-1)^{M(n+1)}    \D H_{\mu }^{(M)}, \label{Torsionm-1}  \\
    E_{\mu }^{(M-1)} &=&  \frac{\partial \mathcal{L}_G}{\partial \D^{M-1} \star T^{\mu }} + (-1)^{(M-1)(n+1)} \star I_{\mu }^{(M-1)}, \nonumber \\
    \label{TorsionHE2}
\end{eqnarray}
which propagate recursively to lower derivative orders; the lowest order being
\begin{eqnarray}
    I_\mu^{(0)} &=& \frac{\partial \mathcal{L}_G}{\partial \star T^{\mu}}  -(-1)^{n} \D E_{\mu}^{(1)},  \label{IGTTH0}  \\
    E_\mu^{(0)} &=& \frac{\partial \mathcal{L}_G}{\partial T^{\mu}} + \star \frac{\partial \mathcal{L}_G}{\partial \star T^{\mu}} -(-1)^{n} \star \D E_\mu^{(1)}.
    \label{EGTNH0}
\end{eqnarray}

Regrouping the variations, a generic action variation takes the form
\begin{equation}
    \delta S = \int \ \delta e^{\mu} \wedge \mathcal{E}_\mu + \delta \omega^{\mu \nu} \wedge\mathcal{E}_{\mu \nu} + \de\theta_{G} (\delta),
    \label{GVtheL}
\end{equation}
where
\begin{widetext}
\begin{eqnarray}
    \mathcal{E}_\mu &=& \frac{\partial \mathcal{L}_G}{\partial e^{\mu}} + \D E_\mu^{(0)}- \text{i}_{\mu} R^{\alpha \beta} \wedge \star I^{(0)}_{\alpha \beta} +\text{i}_{\mu} \star R^{\alpha \beta} \wedge I^{(0)}_{\alpha \beta} + \sum_{j=1}^{N} \left[ \text{i}_\mu \star \D^{j} \star R^{\alpha \beta} \wedge I^{(j)}_{\alpha \beta} - (-1)^{j(n+1)}\text{i}_\mu  \D^{j} \star R^{\alpha \beta} \wedge \star I^{(j)}_{\alpha \beta}\right] \nonumber \\
    && - \text{i}_{\mu} T^{\alpha } \wedge \star I^{(0)}_{\alpha} + \text{i}_{\mu} \star T^{\alpha} \wedge I^{(0)}_\alpha + \sum_{j=1}^{M} \left[ \text{i}_\mu \star \D^{j} \star T^{\alpha } \wedge I^{(j)}_{\alpha } - (-1)^{j(n+1)}\text{i}_\mu  \D^{j} \star T^{\alpha } \wedge \star I^{(j)}_{\alpha }\right],\label{EomemGV}\\
  \mathcal{E}_{\mu \nu} &=& \D H_{\mu \nu}^{(0)}   - e_{[\mu} \wedge E_{\nu]}^{(0)}- \star T_{[\mu} \wedge E_{\nu]}^{(1)}- 2 \star R_{[\mu}^{\ \ \ \alpha} \wedge H_{\nu] \alpha}^{(1)}- \sum_{i=1}^{N-1} 2 \star \D^{i} \star R_{[\mu}^{\ \ \ \alpha} \wedge H_{\nu] \alpha}^{(i+1)}  - \sum_{i=1}^{M-1} \star \D^{i} \star T_{[\mu} \wedge E_{\nu]}^{(i+1)},\nonumber\\&&\label{hodeomA4} \\ 
 \theta_{G}(\delta) &=& \delta \omega^{\mu \nu} \wedge H_{\mu \nu}^{(0)}+ \delta e^\mu \wedge E_{\mu}^{(0)} + \delta \star R^{\mu \nu} \wedge H_{\mu \nu}^{(1)}+ \delta \star T^{\mu} \wedge E_{\mu}^{(1)} +\sum_{i=1}^{N-1} \delta \star \D^{i} \star R^{\mu \nu} \wedge H_{\mu \nu}^{(i+1)} \nonumber \\
 &&+ \sum_{i=1}^{M-1}\delta \star \D^{i} \star T^{\mu} \wedge E_{\mu}^{(i+1)}.
 \label{BTTHeL}
\end{eqnarray}
\end{widetext}
Clearly, the equations of motion are $\mathcal{E}_\mu=0$ and  $\mathcal{E}_{\mu \nu}=0$.

At this stage, the Noether charges associated with diffeomorphisms and GT can be computed. In turn, these objects provide the necessary ingredients to construct the gauge-invariant boundary term and charge, whose explicit evaluation constitutes the main objective of the following subsection.

\subsection{Gauge-invariant boundary term and charge}

The transformations of the vielbein and the spin connection under diffeomorphisms and gauge transformations are respectively given by
\begin{eqnarray}\label{Diffe}
\delta_{\text{Diff}}(\xi) e^{\mu}
&=&  \mathcal{L}_{\xi} e^{\mu} , \\
 \delta_{\text{Diff}}(\xi) \omega^{\mu \nu} &=& \mathcal{L}_{\xi} \omega^{\mu \nu},
\label{DiffeCone}\\
\label{gauge}
 \delta_{\text{GT}}(\lambda) e^\mu &=& - \lambda^\mu_{\ \nu} e^\nu,\\
 \delta_{\text{GT}}(\lambda) \omega^\mu_{\ \nu} &=& \D \lambda^\mu_{\ \nu},
      \end{eqnarray}
With these transformations one can show that 
\begin{eqnarray}
    \theta_{G}[\delta_{\text{GT}}(\lambda)]  &=& \D \lambda^{\mu \nu} \wedge H_{\mu \nu}^{(0)} -\lambda^\mu_{\ \nu} e^\nu \wedge E_{\mu}^{(0)} \nonumber \\
 &&- \lambda^{\mu}_{\ \alpha} \star R^{\alpha \nu} \wedge H_{\mu \nu}^{(1)}- \lambda^{\nu}_{\ \alpha} \star R^{\mu \alpha} \wedge H_{\mu \nu}^{(1)} \nonumber \\
 &&- \lambda^\mu_{\ \alpha} \star T^{\alpha} \wedge E_{\mu}^{(1)} \nonumber \\
 && -\sum_{i=1}^{N-1} \lambda^\mu_{\ \alpha} \star \D^{i} \star R^{\alpha \nu} \wedge H_{\mu \nu}^{(i+1)} \nonumber \\
  && -\sum_{i=1}^{N-1} \lambda^\nu_{\ \alpha} \star \D^{i} \star R^{\mu \alpha} \wedge H_{\mu \nu}^{(i+1)} \nonumber \\
 &&- \sum_{i=1}^{M-1}\lambda^\mu_{\ \nu} \star \D^{i} \star T^{\nu} \wedge E_{\mu}^{(i+1)},
    \label{thetaGt}
\end{eqnarray}
which, after straightforward manipulations, can be rewritten as
\begin{eqnarray}
    \theta_{G}[\delta_{\text{GT}}(\lambda)]  = \de \left[ \lambda^{\mu \nu} H_{\mu \nu}^{(0)}\right] - \lambda^{\mu \nu} \mathcal{E}_{\mu \nu}.
    \label{thetaGT2}
\end{eqnarray}
Notice that the second term vanishes on shell. Consequently, the Noether charge associated with a gauge transformation is 
\begin{equation}
    Q_{\text{GT}} (\lambda) = \lambda^{\mu \nu} H_{\mu \nu}^{(0)}.
    \label{GTchargelambda}
\end{equation}
Moreover, one can propose
\begin{equation}
     \gamma(\delta) = -e^\mu_a \delta e^{\nu a} H_{\mu \nu}^{(0)},
    \label{gammaQGT}
\end{equation}
which clearly satisfies $\gamma[\delta_{\text{GT}}(\lambda)] = - Q_{\text{GT}}(\lambda)$.

On the other hand, it is straightforward to obtain conservation rules associated with gauge symmetry~\cite{BonCris}. Since the action is gauge invariant, the variation satisfies
\begin{eqnarray}
0 &=& \int \lambda^{\mu \nu} \left( e_{\mu} \wedge \mathcal{E}_{\nu} - \text{D} \mathcal{E}_{\mu \nu} \right),
\label{GTvarS}
\end{eqnarray}
for any gauge parameter $\lambda^{\mu \nu} $. Note that, Eq.~\eqref{thetaGT2} is used in the last equation. Hence, demanding invariance under arbitrary gauge transformations implies
\begin{equation}
\text{D} \mathcal{E}_{\mu \nu} = e_{[\mu} \wedge \mathcal{E}_{\nu]}.
\label{Deom}
\end{equation}

The conservation law associated with diffeomorphism invariance can be derived through an analogous procedure. In particular, the variation of the action under an infinitesimal diffeomorphism generated by an arbitrary vector field $\xi$ takes the form
\begin{eqnarray}
\delta_{\text{Diff}}(\xi) S 
 &=& \int \ \xi^{\mu}\Big[-\D \mathcal{E}_{\mu}
 +\text{i}_{\mu} T^{\alpha} \wedge \mathcal{E}_{\alpha}
 +\text{i}_{\mu} R^{\alpha \beta} \wedge \mathcal{E}_{\alpha \beta}\Big] \nonumber \\
 &&+ \de \left[ \xi^{\mu} \mathcal{E}_{\mu}
 + \text{i}_{\xi} \omega^{\mu \nu} \mathcal{E}_{\mu \nu}
 +\theta( \mathcal{L}_{\xi} ) \right],
\label{Lieder}
\end{eqnarray}
where Eq.~\eqref{Deom} is used together with the identities
\begin{eqnarray}\label{Diffe1}
\mathcal{L}_{\xi} e^{\mu} 
&=& \D\,\text{i}_\xi e^{\mu}
 + \text{i}_\xi \D e^\mu
 - \text{i}_{\xi} \omega^\mu_{\ \nu} e^{\nu}, \\
\mathcal{L}_{\xi} \omega^{\mu \nu}
&=& \text{i}_\xi R^{\mu \nu}
 + \D\,\text{i}_{\xi}  \omega^{\mu \nu}.
\label{Diffc}
\end{eqnarray}
This requires
\begin{equation}
\D \mathcal{E}_{\mu}
= \text{i}_{\mu} T^{\alpha} \wedge \mathcal{E}_{\alpha}
+ \text{i}_{\mu} R^{\alpha \beta} \wedge \mathcal{E}_{\alpha \beta},
\label{Bianchieoms}
\end{equation}
to ensure the vanishing of the bulk term in Eq.~\eqref{Lieder}. What is more, since diffeomorphisms are pseudo-symmetries (i.e., invariant up to a boundary term), the variation of the action under an infinitesimal diffeomorphism must reduce to a boundary contribution,
\begin{equation}\label{Lieder2}
    \delta_{\text{Diff}} (\xi) S
    = \int \de \, \mathrm{i}_{\xi} \mathcal{L}_G .
\end{equation}
By comparing the boundary terms in Eqs.~\eqref{Lieder} and \eqref{Lieder2}, one obtains
\begin{eqnarray}
 \mathrm{i}_{\xi} \mathcal{L}_G &=&  \xi^{\mu} \mathcal{E}_{\mu}
 + \mathrm{i}_{\xi} \omega^{\mu \nu} \mathcal{E}_{\mu \nu}
 + \theta_{G}( \mathcal{L}_{\xi} ) + \de \alpha ,
\label{interiorL0}
\end{eqnarray}
where $\alpha$ is an $(n{-}2)$-form resulting from the ambiguity associated with the addition of an exact term. This ambiguity can be fixed by requiring
\begin{equation}
    \theta_{G}(\mathcal{L}_{f \xi}) + \mathrm{d}\alpha(f \xi)
    = f\,[\,\theta_{G}(\mathcal{L}_{\xi}) + \mathrm{d}\alpha(\xi)\,],
    \label{ConditionCharge}
\end{equation}
for any function $f$. Note that
\begin{eqnarray}
 \theta_{G}(\mathcal{L}_{f\xi}) &=& \mathcal{L}_{f\xi} \omega^{\mu \nu} \wedge H_{\mu \nu}^{(0)}+ \mathcal{L}_{f\xi} e^\mu \wedge E_{\mu}^{(0)} \nonumber \\
 &&+ \mathcal{L}_{f\xi} \star R^{\mu \nu} \wedge H_{\mu \nu}^{(1)}+ \mathcal{L}_{f\xi} \star T^{\mu} \wedge E_{\mu}^{(1)} \nonumber \\
 && +\sum_{i=1}^{N-1} \mathcal{L}_{f\xi} \star \D^{i} \star R^{\mu \nu} \wedge H_{\mu \nu}^{(i+1)} \nonumber \\
 &&+ \sum_{i=1}^{M-1}\mathcal{L}_{f\xi} \star \D^{i} \star T^{\mu} \wedge E_{\mu}^{(i+1)}.
 \label{fxitheta}
\end{eqnarray}
In addition, using $\mathcal{L}_{f\xi} \alpha = f \mathcal{L}_{\xi}\alpha + \de f \wedge \mathrm{i}_{\xi}\alpha$, which follows from Cartan’s magic formula, one gets
\begin{eqnarray}
    \mathcal{L}_{f\xi} \omega^{\mu \nu} \wedge H_{\mu \nu}^{(0)} 
    &=& f \left[ \mathcal{L}_\xi \omega^{\mu \nu} \wedge H_{\mu \nu}^{(0)}\right] 
    + \de f \wedge \text{i}_\xi \omega^{\mu \nu} H_{\mu \nu}^{(0)} \nonumber \\
    &=& f \left[ \mathcal{L}_\xi \omega^{\mu \nu} \wedge H_{\mu \nu}^{(0)}\right] 
    - f\, \D \text{i}_\xi \omega^{\mu \nu} \wedge H_{\mu \nu}^{(0)} \nonumber \\
    && - f\, \text{i}_\xi \omega^{\mu \nu} \D H_{\mu \nu}^{(0)} 
    + \de \left[ f\, \text{i}_\xi \omega^{\mu \nu} H_{\mu \nu}^{(0)}\right] \nonumber \\
    &=&  f \left[ \text{i}_\xi R^{\mu \nu} \wedge H_{\mu \nu}^{(0)}\right] 
    - f\, \text{i}_\xi \omega^{\mu \nu} \D H_{\mu \nu}^{(0)} 
    \nonumber \\
    &&+ \de \left[ f\, \text{i}_\xi \omega^{\mu \nu} H_{\mu \nu}^{(0)}\right].
    \label{connectionlie}
\end{eqnarray}

Performing analogous manipulations for the second term in Eq.~\eqref{fxitheta} yields
\begin{eqnarray}
    \mathcal{L}_{f\xi} e^\mu \wedge E_{\mu}^{(0)}  
    &=& f \left[ \text{i}_\xi T^{\mu} \wedge E_{\mu}^{(0)} 
    + \text{i}_\xi \omega^{\mu \nu} e_{[\mu} \wedge E_{\nu]}^{(0)} \right] \nonumber \\
    && - f\, \xi^{\mu} \D E_{\mu}^{(0)} 
    + \de \left[f\, \xi^{\mu} E_{\mu}^{(0)} \right].
    \label{tetradLief}
\end{eqnarray}
The only terms in which there are derivatives of $f$ are exact terms in Eqs.~\eqref{connectionlie} and \eqref{tetradLief}, which can be canceled by fixing
\begin{eqnarray}
   - \alpha &=& \xi^\mu E_\mu^{(0)} + \text{i}_\xi \omega^{\mu \nu} H_{\mu \nu}^{(0)} \nonumber \\ 
   && + \text{i}_\xi \star R^{\mu \nu} \wedge H_{\mu \nu}^{(1)} 
   + \text{i}_\xi \star T^\mu \wedge E_\mu^{(1)} \nonumber \\
   && + \sum_{i=1}^{N-1} \text{i}_{\xi} \star \D^{i} \star R^{\mu \nu} \wedge H_{\mu \nu}^{(i+1)} \nonumber \\
   && + \sum_{i=1}^{M-1} \text{i}_{\xi} \star \D^{i} \star T^{\mu} \wedge E_{\mu}^{(i+1)} .
    \label{alphafix}
\end{eqnarray}

What is more, from Eqs.~\eqref{Jo} and \eqref{interiorL0} it is clear that 
\begin{eqnarray}
 J(\xi) &=& \theta_{G}(\mathcal{L}_{\xi}) - \text{i}_\xi \mathcal{L}_G \nonumber \\
 &=& -\text{i}_{\xi}e^{\mu} \mathcal{E}_{\mu} - \text{i}_{\xi} \omega^{\mu \nu} \mathcal{E}_{\mu \nu} - \de \alpha (\xi),
 \label{NCurrentGV}
\end{eqnarray}
which is conserved on shell. Thus, its associated Noether charge is
\begin{equation}
 Q_{G}(\xi) = -\alpha(\xi) 
 \label{NoetherchargeGV}
\end{equation}
where other ambiguities are ignored.

With all this, the gauge-invariant Noether charge can be computed. It has the form 
\begin{eqnarray}
    \hat{Q}_{G}(\xi) &=& Q_{G}(\xi) - Q_\text{GT} (\lambda_\xi) \nonumber \\
    &=& \xi^\mu E_\mu^{(0)} -\mathring{\nabla}^\mu \xi^{\nu} H_{\mu \nu}^{(0)}+ \text{i}_\xi K^{\mu \nu} H_{\mu \nu}^{(0)} \nonumber \\ 
   && + \text{i}_\xi \star R^{\mu \nu} \wedge H_{\mu \nu}^{(1)} 
   + \text{i}_\xi \star T^\mu \wedge E_\mu^{(1)} \nonumber \\
   && + \sum_{i=1}^{N-1} \text{i}_{\xi} \star \D^{i} \star R^{\mu \nu} \wedge H_{\mu \nu}^{(i+1)} \nonumber \\
   && + \sum_{i=1}^{M-1} \text{i}_{\xi} \star \D^{i} \star T^{\mu} \wedge E_{\mu}^{(i+1)},
    \label{Qhat}
\end{eqnarray}
where Eq.~\eqref{decomposicion} is used. In addition, the corrected boundary term is
\begin{eqnarray}
   \hat{\theta}_{G}(\delta) &=& \theta_{G}(\delta) + \de \gamma(\delta) \nonumber \\ 
   &=&\delta \omega^{\mu \nu} \wedge H_{\mu \nu}^{(0)}+ \delta e^\mu \wedge E_{\mu}^{(0)} \nonumber \\
 &&+ \delta \star R^{\mu \nu} \wedge H_{\mu \nu}^{(1)}+ \delta \star T^{\mu} \wedge E_{\mu}^{(1)} \nonumber \\
 && +\sum_{i=1}^{N-1} \delta \star \D^{i} \star R^{\mu \nu} \wedge H_{\mu \nu}^{(i+1)} \nonumber \\
 &&+ \sum_{i=1}^{M-1}\delta \star \D^{i} \star T^{\mu} \wedge E_{\mu}^{(i+1)} \nonumber \\
 &&- \de \left[ e^{\mu}_a \delta e^{\nu a} H_{\mu \nu}^{(0)}\right].
 \label{BTcorrecGV}
\end{eqnarray}

At this stage, the first law of black hole thermodynamics can be computed using Eq.~\eqref{1stlawcorrected}. It can be written as
\begin{equation}
    \int_{\mathcal{B}} \delta \left[ -\nabla^\mu \xi^\nu H_{\mu \nu}^{(0)}\right] = \int_{\text{i}^0} \delta Q_{G} (\xi) - \text{i}_\xi \theta_G (\delta),
    \label{GravitationalCharge}
\end{equation}
where $\mathcal{B}$ is a co-dimension 2 surface. Notice that, since $H_{\mu \nu}^{(0)}$ is an $(n-2)$-form, being obtained from the variation of the Lagrangian top-form with respect to the curvature 2-form, the integrand on the left-hand side is also an $(n-2)$-form. Therefore, the integral in Eq.~\eqref{GravitationalCharge} is well defined. The next task is to incorporate matter fields in this formalism, which is done in the next section.

\section{Matter Fields}\label{MFie}

The main goal of this part of the paper is to incorporate matter fields into the framework developed in previous sections. Matter is represented by differential $p$-forms and can be divided into two types: ordinary fields (such as scalar fields, spinors, etc.), which are collectively denoted by $\Phi^{I}$, and gauge fields (connections), denoted by $\mathit{A}^{I}_{\ J}$, which are 1-forms taking values in a Lie algebra $\mathfrak{g}$. Note that capital Latin indices are associated with $\mathfrak{g}$.

For the case at hand, the structure group associated with a gravitational field coupled to a gauge field is $SO(n-1,1)\times G$, where $G$ is the Lie group associated with gauge invariance, whose whose Lie algebra is $\mathfrak{g}$. Hence, the algebra of the theory is $\mathfrak{so}(n-1,1)\oplus\mathfrak{g}$. Then, the index $I$ labels the representation space on which this Lie algebra acts. Therefore, it can be decomposed as $I=(\bar{k},\bar{I})$, where $\bar{k}$ is associated with a representation of the Lorentz algebra and $\bar{I}$ with a representation of $\mathfrak{g}$. In such a representation, the connection 1-form takes the form
\begin{equation}
    \mathcal{A}^{I}_{\ J}
    =
    \delta^{\bar{I}}_{\bar{J}} \,
    \omega^{\bar{i}}_{\ \bar{j}}
    +
    A^{\bar{I}}_{\ \bar{J}}
    \delta^{\bar{i}}_{\ \bar{j}},
    \label{ConnectionA}
\end{equation}
where
\begin{equation}
    \omega^{\bar{k}}_{\ \bar{l}}
    =
    \frac{1}{2}
    \omega_{\mu \nu}
    (\mathfrak{J}^{\mu \nu})^{\bar{k}}_{\ \bar{l}},
\end{equation}
and $(\mathfrak{J}^{\mu \nu})^{\bar{k}}_{\ \bar{l}}$ are the generators of the Lorentz algebra in the corresponding representation.

By definition, the curvature 2-form associated with Eq.~\eqref{ConnectionA} is 
\begin{eqnarray}
    \mathcal{R}^{I}_{\ J}
    &=&
    \de \mathcal{A}^{I}_{\ J}
    +
    \mathcal{A}^{I}_{\ K}
    \wedge
    \mathcal{A}^{K}_{\ J}
    \nonumber\\
    &=&
    \delta^{\bar{I}}_{\bar{J}}
    \de \omega^{\bar{i}}_{\ \bar{j}}
    +
    \left[
    \de A^{\bar{I}}_{\ \bar{J}}
    \right]
    \delta^{\bar{i}}_{\ \bar{j}}
    \nonumber\\
    &&
    +
    \left[
    \delta^{\bar{I}}_{\bar{K}}
    \omega^{\bar{i}}_{\ \bar{k}}
    +
    A^{\bar{I}}_{\ \bar{K}}
    \delta^{\bar{i}}_{\ \bar{k}}
    \right]
    \wedge
    \left[
    \delta^{\bar{K}}_{\bar{J}}
    \omega^{\bar{k}}_{\ \bar{j}}
    +
    A^{\bar{K}}_{\ \bar{J}}
    \delta^{\bar{k}}_{\ \bar{j}}
    \right]
    \nonumber\\
    &=&
    \delta^{\bar{I}}_{\bar{J}}
    \left[
    \de \omega^{\bar{i}}_{\ \bar{j}}
    +
    \omega^{\bar{i}}_{\ \bar{k}}
    \wedge
    \omega^{\bar{k}}_{\ \bar{j}}
    \right]
    \nonumber\\
    &&
    +
    \left[
    \de A^{\bar{I}}_{\ \bar{J}}
    +
    A^{\bar{I}}_{\ \bar{K}}
    \wedge
    A^{\bar{K}}_{\ \bar{J}}
    \right]
    \delta^{\bar{i}}_{\ \bar{j}}
    \nonumber\\
    &=&
    \delta^{\bar{I}}_{\bar{J}}
    \left[
    R^{\bar{i}}_{\ \bar{j}}
    \right]
    +
    \left[
    F^{\bar{I}}_{\ \bar{J}}
    \right]
    \delta^{\bar{i}}_{\ \bar{j}}.
    \label{CurvaTotal}
\end{eqnarray}
What is more,
\begin{eqnarray}
    R^{\bar{i}}_{\ \bar{j}}
    &=&
    \frac{1}{2}
    R_{\mu \nu}
    (\mathfrak{J}^{\mu \nu})^{\bar{i}}_{\ \bar{j}}
        =
    \de \omega^{\bar{i}}_{\ \bar{j}}
    +
    \omega^{\bar{i}}_{\ \bar{k}}
    \wedge
    \omega^{\bar{k}}_{\ \bar{j}},
\end{eqnarray}
and
\begin{equation}
    F^{\bar{I}}_{\ \bar{J}}
    =
    \de A^{\bar{I}}_{\ \bar{J}}
    +
    A^{\bar{I}}_{\ \bar{K}}
    \wedge
    A^{\bar{K}}_{\ \bar{J}},
\end{equation}
are the curvature 2-forms associated with the Lorentz and internal gauge fields, respectively.

It is worth noting that, for a tensorial representation,
\begin{equation}
\left( \mathfrak{J}^{\mu \nu}\right)^\alpha_{\ \beta}
=
\eta^{\mu \alpha} \delta^{\nu}_\beta
-
\eta^{\nu \alpha} \delta^{\mu}_\beta.
\end{equation}
In this case, Eq.~\eqref{ConnectionA} reproduces the standard Lorentz connection $\omega^\alpha{}_\beta$, while Eq.~\eqref{CurvaTotal} yields the curvature 2-form $R^\alpha{}_\beta$ given in Eq.~\eqref{2curvaform}. For simplicity, the notation
\begin{eqnarray}
    F^{I}_{\ J}
    &=&
    \delta^{\bar{i}}_{\ \bar{j}}
    F^{\bar{I}}_{\ \bar{J}}, \label{curvintGauge}\\
    R^{I}_{\ J}
    &=&
    R^{\bar{i}}_{\ \bar{j}}
    \delta^{\bar{I}}_{\ \bar{J}},
\end{eqnarray}
is adopted.

The gauge-covariant derivative associated with the internal connection
$A^{\bar I}{}_{\bar J}$ acts on a $\mathfrak g$-valued $p$-form
$U^{\bar I_1 \dots \bar I_p}{}_{\bar J_1 \dots \bar J_q}$ as
\begin{eqnarray}
\bar{\D} U^{\bar I_1 \dots \bar I_p}_{\ \ \ \ \ \ \bar J_1 \dots \bar J_q}
&=&
\de U^{\bar I_1 \dots \bar I_p}_{\ \ \ \ \ \ \bar J_1 \dots \bar J_q}
+
A^{\bar I_1}_{\ \bar K}
\wedge
U^{\bar K \dots \bar I_p}_{\ \ \ \ \ \ \bar J_1 \dots \bar J_q}
\nonumber\\
&&
+\dots
-
A^{\bar K}_{\ \bar J_q}
\wedge
U^{\bar I_1 \dots \bar I_p}_{\ \ \ \ \ \ \bar J_1 \dots \bar K}.
\label{covDFA}
\end{eqnarray}
Then, the Bianchi identities take the form
\begin{eqnarray}
    \D R^{I}_{\ J}
    &=&
    \frac{1}{2}\D R_{\mu\nu}
    \left(\mathfrak{J}^{\mu\nu}\right)^{\bar i}_{\ \bar j} \delta^{\bar I}_{\ \bar J}
    =
    0,
    \\
    \D F^{I}_{\ J}
    &=&
    \delta^{\bar i}_{\ \bar j}
    \bar{\D} F^{\bar I}_{\ \bar J}
    =
    0.
    \label{Bianchies}
\end{eqnarray}

Matter fields may transform nontrivially under both, the internal gauge group and the Lorentz group. In this case, the corresponding covariant derivative is
\begin{equation}
\D \Phi^{I} = \de \Phi^{I} + \mathit{A}^{I}_{\ J} \wedge \Phi^{J} + \frac{1}{2} \omega_{\mu \nu} (\mathfrak{J}^{\mu \nu})^{I}_{\ J} \wedge \Phi^{J}.
\label{FullCovD}
\end{equation}
Clearly, for matter fields carrying a trivial Lorentz representation, the last term vanishes and Eq.~\eqref{FullCovD} reduces to the gauge-covariant derivative associated with the internal connection.

For example, for a spinor field carrying no gauge charge, the covariant derivative reduces to
\begin{equation}
\D \Phi^{\bar i}
=
\de \Phi^{\bar i}
+
\frac{1}{2}
\omega_{\mu \nu}
\left(\Sigma^{\mu \nu}\right)^{\bar i}{}_{\bar j}
\Phi^{\bar j},
\label{covdspin}
\end{equation}
where
\begin{equation}
\Sigma^{\mu\nu}
=
\frac{i}{4}
[\gamma^\mu,\gamma^\nu]
\end{equation}
are the generators of the Lorentz algebra in the spinor representation, which are given in terms of the Dirac matrices satisfying
\begin{equation}
\{\gamma^\mu,\gamma^\nu\}
=
-2\eta^{\mu\nu}.
\end{equation}

The generic matter Lagrangian, which includes nonminimal couplings, can be written as
\begin{eqnarray}
\mathcal{L}_M = \mathcal{L}_M [ && e, R,\D \star R ,\star \D \star R, \dots , \D^{N} \star R, \star \D^{N} \star R  ; \nonumber \\
&&\Phi, \star \Phi, \D \Phi, \D \star \Phi, \dots, \D^{k} \Phi, \D^{l}\star \Phi, \nonumber \\ && \star\D \Phi, \star\D \star \Phi, \dots , \star \D^{k} \Phi, \star \D^{l}\star \Phi, \nonumber \\ &&\mathit{F}, \star \mathit{F}, \D \star \mathit{F} , \star \D \star F, \dots, \D^{h} \star F, \star \D^{h} \star F], \nonumber \\ \label{Matter}
\end{eqnarray}
where $k$, $l$, and $h$ denote the maximum order of derivatives, and the convention for $\D^{i}$ is generalized from that used in the gravity sector. Note that, in view of the decomposition given in Eq.~\eqref{curvintGauge}, the action of $\D$ on $F^{I}{}_{J}$ is insensitive to the Lorentz connection. Hence, terms involving $\star F^{I}{}_{J}$ depend only on the internal gauge connection.

The variation of the matter Lagrangian yields the matter equations of motion. It also yields the energy-momentum tensor, $\tau_\mu$, and the spin density, $S_{\mu \nu}$, given be the variations of $\mathcal{L}_{M}$ with respect to $e^{\mu}$ and $\omega^{\mu \nu}$, respectively. In fact, a general variation of $\mathcal{L}_{M}$ can be written as
\begin{widetext}
\begin{eqnarray}
\delta \mathcal{L}_{M} &=& \delta e^{\mu} \wedge \frac{\partial \mathcal{L}_M}{\partial e^{\mu}} + \delta R^{\mu \nu} \wedge \frac{\partial \mathcal{L}_M}{\partial R^{\mu \nu}} + \delta T^{\mu} \wedge \frac{\partial \mathcal{L}_M}{\partial T^{\mu}}  +\delta \star R^{\mu \nu} \wedge \frac{\partial \mathcal{L}_M}{\partial \star R^{\mu \nu}} + \delta \star T^{\mu} \wedge \frac{\partial \mathcal{L}_M}{\partial \star T^{\mu}} \nonumber \\
    && \ + \sum_{j=1}^{N} \delta \star \D^{j}  \star R^{\mu \nu} \wedge  \frac{\partial \mathcal{L}_M}{\partial \star \D^{j}  \star R^{\mu \nu}} +   \sum_{j=1}^{N} \delta  \D^{j}  \star R^{\mu \nu} \wedge  \frac{\partial \mathcal{L}_M}{\partial \D^{j}  \star R^{\mu \nu}} \nonumber \\
    &&  \ + \sum_{j=1}^{M} \delta \star \D^{j}  \star T^{\mu} \wedge  \frac{\partial \mathcal{L}_M}{\partial \star \D^{j}  \star T^{\mu }} +   \sum_{j=1}^{M} \delta  \D^{j}  \star T^{\mu} \wedge  \frac{\partial \mathcal{L}_M}{\partial \D^{j}  \star T^{\mu}} \nonumber \\
    &&
+\delta \Phi^{I} \wedge \frac{\partial\mathcal{L}_{M}}{\partial \Phi^{I}} + \delta \star \Phi^{I} \wedge \frac{\partial L_{M}}{\partial \star \Phi^{I}}  + \delta \D \Phi^{I} \wedge \frac{\partial \mathcal{L}_{M}}{\partial \D\Phi^{I}}\nonumber \\
&&  + \delta \star \D \Phi^{I} \wedge \frac{\partial \mathcal{L}_{M}}{\partial \star \D\Phi^{I}} + \delta \D \star \Phi^{I} \wedge \frac{\partial \mathcal{L}_{M}}{\partial \D \star \Phi^{I}}  +  \delta \star \D \star \Phi^{I} \wedge \frac{\partial \mathcal{L}_{M}}{\partial \bar{\star \D \star}\Phi^{I}} + \sum^{k}_{i=2} \delta \D^{i} \Phi^{I} \wedge \frac{\partial \mathcal{L}_{M}}{\partial \D^{i} \Phi^{I}}\nonumber \\
&&+ \sum^{l}_{i=2} \delta \D^{i} \star \Phi^{I} \wedge \frac{\partial \mathcal{L}_{M}}{\partial \D^{i} \star \Phi^{I}} + \sum^{k}_{i=2} \delta \star \D^{i} \Phi^{I} \wedge \frac{\partial \mathcal{L}_{M}}{\partial \star \D^{i}  \Phi^{I}}  + \sum^{l}_{i=2} \delta \star \D^{i} \star \Phi^{I} \wedge \frac{\partial \mathcal{L}_{M}}{\partial\star \D^{i} \star \Phi^{I}} \nonumber \\
&& + \delta \mathit{F}^{I}_{\ J} \wedge \frac{\partial \mathcal{L}_{M}}{\partial \mathit{F}^{I}_{\ J}} + \delta \star \mathit{F}^{I}_{\ J} \wedge \frac{\partial \mathcal{L}_{M}}{\partial \star \mathit{F}^{I}_{\ J}}  +\sum^{h}_{i=1} \delta \D^{i} \star \mathit{F}^{I}_{\ J} \wedge \frac{\partial \mathcal{L}_{M}}{\partial \D^{i} \star \mathit{F}^{I}_{\ J}} + \sum^{h}_{i=1} \delta \star \D^{i} \star \mathit{F}^{I}_{\ J} \wedge \frac{\partial \mathcal{L}_{M}}{\partial \star \D^{i} \star \mathit{F}^{I}_{\ J}} ,
\label{LMvar}
\end{eqnarray}
Notice that this variation has the same structure as the variation of the gravity Lagrangian given in Eq.~\eqref{TheL}. Still, the explicit computations of the energy-momentum and spin density are presented in Appendix~\ref{EOMGTG}.

The boundary term takes the form
\begin{eqnarray}
 \theta_M(\delta) &=& \delta \omega^{\mu \nu} \wedge \mathcal{H}_{\mu \nu}^{(0)}
+ \delta e^\mu \wedge \mathcal{E}_{\mu}^{(0)} + \delta \star R^{\mu \nu} \wedge \mathcal{H}_{\mu \nu}^{(1)}
+ \delta \star T^{\mu} \wedge \mathcal{E}_{\mu}^{(1)} \nonumber \\
&&+\sum_{i=1}^{N-1}
\delta \star \D^{i} \star R^{\mu \nu}
\wedge \mathcal{H}_{\mu \nu}^{(i+1)}
+ \sum_{i=1}^{M-1}
\delta \star \D^{i} \star T^{\mu}
\wedge \mathcal{E}_{\mu}^{(i+1)}
\nonumber \\
&&+\delta \Phi^{I} \wedge E_{I}^{(1)}
+ \delta \star \Phi^{I} \wedge G_{I}^{(1)}
 + \sum_{j=1}^{k-1}
\delta \star \D^{j} \Phi^{I}
\wedge E_{I}^{(j+1)}
\nonumber \\
&&+ \sum_{j=1}^{l-1}
\delta \star \D^{j} \star \Phi^{I}
\wedge G_{I}^{(j+1)}
 + \delta \mathit{A}^{I}_{\ J}
\wedge [H^{(0)}]_{I}^{\ J}+ \delta \star F^I_{\ J} \wedge  [H^{(1)}]_{I}^{\ J} \nonumber \\
&&
 + \sum_{j=1}^{h-1}
\delta \D^{j}\star \mathit{F}^{I}_{\ J}
\wedge [H^{(j+1)}]_{I}^{\ J},
 \label{BTMatter}
\end{eqnarray}
\end{widetext}
where $\mathcal{E}_{\mu}^{(i)}$ and $\mathcal{H}_{\mu\nu}^{(i)}$ denote the matter counterparts of the gravitational coefficients $E_{\mu}^{(i)}$ and $H_{\mu\nu}^{(i)}$, while $E_{I}^{(j)}$, $G_{I}^{(j)}$ and $[H^{(j)}]_{I}^{\ J}$ are defined in Eqs.~\eqref{Ek-1}, \eqref{starmatter} and \eqref{H1Matter3}, respectively. Hence, a general variation of the matter Lagrangian can be written as
\begin{eqnarray}
    \delta \mathcal{L}_M &=& 
    -\,\delta e^\mu \wedge \tau_\mu 
    - \delta \omega^{\mu \nu} \wedge S_{\mu \nu}  \nonumber \\
    && + \delta \Phi^{I} \wedge \mathcal{E}_I 
    + \delta \mathit{A}^{I}_{\ J} \wedge \mathcal{E}_I^{\ J} 
    + \de \theta_M (\delta),
    \label{VarMatterfields}
\end{eqnarray}
where $\mathcal{E}_I$ denotes the quantity whose vanishing yields the equation of motion for $\Phi^I$ [Eq.~\eqref{EOMMAtterPhi}], and $\mathcal{E}_I^{\ J}$ denotes the analogous quantity associated with $\mathit{A}^I_{\ J}$ [Eq.~\eqref{EOMmatterA}].

The most general theory has a gravity sector and a matter sector. In fact, its action is
\begin{equation}
    S = \int \  \mathcal{L}_G+ \mathcal{L}_M,
    \label{LagrangianTotal}
\end{equation}
where $\mathcal{L}_G$ is given in Eq.~\eqref{TheL} and $\mathcal{L}_M$ in Eq.~\eqref{Matter}.
A general variation of Eq.~\eqref{LagrangianTotal} produces
\begin{eqnarray}
    \delta S &=& \int \Big[
    \delta e^{\mu } \wedge \left( \mathcal{E}_\mu - \tau_\mu\right)
    + \delta \omega^{\mu \nu} \wedge \left( \mathcal{E}_{\mu \nu} - S_{\mu \nu}\right) \nonumber \\
    && \quad +\, \delta \Phi^{I} \wedge \mathcal{E}_I 
    + \delta \mathit{A}^{I}_{\ J} \wedge \mathcal{E}_I^{\ J} 
    + \de \theta (\delta)\Big],
    \label{varLGM}
\end{eqnarray}
where $\theta = \theta_G + \theta_M$, with $\theta_G$ is given in Eq.~\eqref{BTTHeL}.

It is important to emphasize that the charges associated with the Lorentz group originate exclusively from the gravitational sector as, by assumption, the matter fields do not carry free Lorentz indices. Hence, only charges associated with diffeomorphism acquire contributions due to the coupling to matter.

On the other hand, diffeomorphism and gauge invariance yield the well known ``conservation'' laws for the energy--momentum tensor and the spin density~\cite{CristobalYuri3,Ruso}. If the matter Lagrangian $\mathcal{L}_{M}$ is invariant under both, diffeomorphisms and internal gauge transformations, the argument developed for the gravitational sector can be directly generalized leading to
\begin{eqnarray}
 \int \de\, \text{i}_{\xi} \mathcal{L} 
 &=& \int \de \bigg[ 
 \text{i}_{\xi} e^{\mu}(\mathcal{E}_\mu- \tau_{\mu}) 
 + \text{i}_{\xi} \omega^{\mu \nu} (\mathcal{E}_{\mu \nu} - S_{\mu \nu})  \nonumber\\
 && \quad +\, \text{i}_{\xi} \Phi^{I} \wedge \mathcal{E}_{I} 
 + \text{i}_{\xi} \mathit{A}^{I}_{\ J} \mathcal{E}_{I}^{\ J} 
 + \theta (\mathcal{L}_{\xi}) \bigg].
 \label{MatteriL}
\end{eqnarray}
This implies that
\begin{eqnarray}
    \text{i}_\xi \mathcal{L} &=& \text{i}_{\xi} e^{\mu}(\mathcal{E}_\mu- \tau_{\mu}) 
    + \text{i}_{\xi} \omega^{\mu \nu} (\mathcal{E}_{\mu \nu} - S_{\mu \nu}) \nonumber \\
    && +\, \text{i}_{\xi} \Phi^{I} \wedge \mathcal{E}_{I} 
    + \text{i}_{\xi} \mathit{A}^{I}_{\ J} \mathcal{E}_{I}^{\ J} 
    + \theta (\mathcal{L}_{\xi}) - \de Q (\xi), \nonumber \\
    &&
    \label{Ltexti}
\end{eqnarray}
where $Q(\xi)$ is fixed by requiring that, under the replacement $\xi \to f\xi$, both sides of Eq.~\eqref{Ltexti} transform in the same way. Consequently, the charge associated with diffeomorphisms splits as
\begin{equation}
    Q(\xi) = Q_G (\xi) + Q_M (\xi),
    \label{Noetherchargemat}
\end{equation}
where the contribution from the matter fields is
\begin{eqnarray}
        Q_M(\xi) &=& \xi^\mu \mathcal{E}_\mu^{(0)} + \text{i}_\xi \omega^{\mu \nu} \mathcal{H}_{\mu \nu}^{(0)} \nonumber \\ 
   && + \text{i}_\xi \star R^{\mu \nu} \wedge \mathcal{H}_{\mu \nu}^{(1)} 
   + \text{i}_\xi \star T^\mu \wedge \mathcal{E}_\mu^{(1)} \nonumber \\
   && + \sum_{i=1}^{N-1} \text{i}_{\xi} \star \D^{i} \star R^{\mu \nu} \wedge \mathcal{H}_{\mu \nu}^{(i+1)} \nonumber \\
   && + \sum_{i=1}^{M-1} \text{i}_{\xi} \star \D^{i} \star T^{\mu} \wedge \mathcal{E}_{\mu}^{(i+1)} \nonumber \\
     && +   \text{i}_\xi \Phi^{I} \wedge E_{I}^{(1)} 
        + \text{i}_\xi \star \Phi^{I} \wedge G_{I}^{(1)}  \nonumber \\
        &&+ \sum_{j=1}^{k-1} \text{i}_{\xi} \star \D^{j} \Phi^{I} \wedge E_{I}^{(j+1)} \nonumber \\
        &&+ \sum_{j=1}^{l-1} \text{i}_{\xi} \star \D^{j} \star \Phi^{I} \wedge H_{I}^{(j+1)} \nonumber \\ 
        && + \text{i}_\xi \mathit{A}^{I}_{\ J} \wedge [H^{(0)}]_{I}^{\ J} 
        + \text{i}_\xi \star \mathit{F}^{I}_{\ J} \wedge [H^{(1)}]_{I}^{\ J} \nonumber \\
        && + \sum_{j=1}^{h-1} \text{i}_{\xi} \star \D^{j} \star \mathit{F}^{I}_{\ J} \wedge [H^{(j+1)}]_{I}^{\ J}.
        \label{NoetherchargeGraMat}
    \end{eqnarray}

Now, the additional contribution to the boundary term and the Noether charges that makes them gauge invariant originates solely from the gravitational sector, as it associated with Lorentz indices. Thus, the gauge charge is given by
\begin{equation}
        Q_{GT} (\lambda) = \lambda^{\mu \nu}\left[ H_{\mu \nu}^{(0)} + \mathcal{H}_{\mu \nu}^{(0)}\right] 
    \label{GtchargeMatter}\end{equation}
In addition, the other relevant object in the analysis, $\gamma(\delta)$ becomes,
\begin{equation}
    \gamma(\delta) =-e^{\mu}_a \delta e^{\nu a}\left[ H_{\mu \nu}^{(0)} + \mathcal{H}_{\mu \nu}^{(0)}\right]
\end{equation}
With this objects it is possible to define the gauge invariant charge and boundary term:
\begin{eqnarray}
    \hat{Q}(\xi) &=& \hat{Q}_G (\xi) + \hat{Q}_M (\xi),
    \label{correctedchargesmat}\\
    \hat{\theta}(\delta) &=& \hat{\theta}_G (\delta) + \hat{\theta}_M (\delta).
    \label{correctedBTmat}
\end{eqnarray}
These quantities are the fundamental ingredients needed in the construction of the first law of black hole thermodynamics. In the next section, these expressions are employed to interpret the entropy and conserved quantities at infinity.


\section{Entropy}\label{Ent}

The central quantities entering the first law of black hole thermodynamics are the gauge invariant Noether charge $\hat{Q}(\xi)$ and boundary term $\hat{\theta}(\delta)$. Moreover, recall that the integration hypersurface $\Sigma$ has two boundaries: one at spatial infinity and the bifurcation surface of the Killing horizon, $\mathcal{B}$, which must be treated separately. On $\mathcal{B}$, the Killing vector field $\xi$ vanishes by definition. Hence,
\begin{equation}
    \int_\mathcal{B} \delta \hat{Q} (\xi) 
    = \int_\mathcal{B} \ \delta \left[- \mathring{\nabla}^\mu \xi^\nu \left( H_{\mu \nu}^{(0)} + \mathcal{H}_{\mu \nu}^{(0)} \right) \right],
    \label{Bifurcationpartxi}
\end{equation}
where $H_{\mu \nu}^{(0)}$ is defined in Eq.~\eqref{Hzero} and $\mathcal{H}_{\mu \nu}^{(0)}$ is obtained through the same construction applied to the matter sector. Recall that these quantities are $(n-2)$-forms. Consequently, the variation of the black hole entropy can be expressed as
\begin{equation}
 \chi \, \delta \!\left(\int_\mathcal{B} n^{\mu \nu}\left[ H_{\mu \nu}^{(0)} + \mathcal{H}_{\mu \nu}^{(0)}\right] \right),
\label{Entropy}
\end{equation}
where $n^{\mu \nu}$ is the binormal to the bifurcation surface, i.e. the antisymmetric tensor constructed from the two independent normal directions to $\mathcal{B}$.

For example, in the Einstein--Hilbert theory, 
\begin{equation}
    H_{\mu \nu}^{(0)}
    =
    \frac{1}{2}\star(e_\mu \wedge e_\nu) ,
    \label{EHcontrib}
\end{equation}
so that
\begin{eqnarray}
    n^{\mu \nu} H_{\mu \nu}^{(0)}
    &=&  
    \frac{1}{2}n^{\mu \nu}\star(e_\mu \wedge e_\nu)=
    \star n  ,
    \label{nbiform}
\end{eqnarray}
where $n= n_{\mu \nu} e^\mu\wedge e^\nu/2$ is the binormal 2-form. Since $\star n$ is the induced (hyper)volume form on $\mathcal{B}$, the Einstein--Hilbert contribution reproduces the geometric $(n-2)$-dimensional volume (or area) in the black-hole entropy. As is shown below, the matter contribution $n^{\mu\nu}\mathcal{H}_{\mu\nu}^{(0)}$ vanishes for minimally coupled matter fields and can modify the entropy only through nonminimal couplings to the gravitational sector.

Hence, in a general theory, the black hole entropy takes the form
\begin{equation}
S = 2 \pi \int_\mathcal{B} n^{\mu \nu} \left[ H_{\mu \nu}^{(0)} + \mathcal{H}_{\mu \nu}^{(0)}\right],
\label{Entropy2}
\end{equation}
where no confusion should arise from the fact that the same symbol $S$ is used for the entropy and for the action, as the integration regions are different.

Equation~\eqref{Entropy2} provides a general expression for the entropy in a stationary and asymptotically flat black hole that has a bifurcated Killing horizon. It should be compared with Eq. \eqref{Hzero}, where it is possible to verify that the entropy arises solely from the Lagrangian terms that contain the curvature two-form. However, curvature includes torsional contributions. Hence, torsion does affect the black hole entropy, but only through curvature. After determining the entropy contribution from the bifurcation surface, it remains to interpret the boundary integral at spatial infinity. This issue is addressed in the next section.

\section{Charges at spatial infinity} \label{HamInv}

To interpret the boundary term at spatial infinity, it is necessary to consider the Hamiltonian formulation. Before doing so, a few remarks are in order. In the metric formalism, the conserved charges at spatial infinity can be obtained from the Hamiltonian associated with the timelike vector field $\xi$, \( H[\xi] \), which satisfies~\cite{MR359663,WaldB,Iyer}  
\begin{equation}
    \delta H[\xi] = \hat{\Omega}(\delta, \mathcal{L}_\xi).
    \label{HamiltonianCI}
\end{equation}
Moreover, for asymptotically flat spacetimes, the Hamiltonian takes the form~\cite{MR359663}
\begin{equation}
    H[\xi] = H_0[\xi] + M_{\mathrm{ADM}},
    \label{Hamiltonian}
\end{equation}
where \( M_{\mathrm{ADM}} \) represents the total energy of spacetime and \( H_0[\xi] \) is constructed from the constraints in such a way that it vanishes  on shell.

The standard construction of the Hamiltonian in diffeomorphism-invariant gravity theories is presented in Refs.~\onlinecite{WaldB,MR359663,PhysRevD.103.064062,PhysRevD.72.104020,CHEN19955}. Nevertheless, since the Hamiltonian variation is determined by the presymplectic current, the conserved charges can be directly obtained from Eq.~\eqref{HamiltonianCI}, once the constraints are satisfied. This approach provides a simpler and equivalent method to construct the Hamiltonian, which is adopted in the remainder of this work. A detailed construction of the relation of presymplectic current and charges at infinity, following the Nester--Gotay construction \cite{CHEN19955}, with which one can show that $\delta H_\xi = \Omega(\delta,\mathcal{L}_\xi)$ for an asymptotically flat solution, is presented in Appendix~\ref{ChargesAtInf}. 

It is essential to clarify which version of the presymplectic current should be employed when evaluating the charges at infinity. From Eq.~\eqref{PreSGT}, it is evident that $\Omega$ and $\hat{\Omega}$ differ by
\begin{equation}
    \Delta \Omega (\delta , \mathcal{L}_\xi) 
    = \int_\Sigma \de \big[ \delta \gamma(\mathcal{L}_\xi) - \mathcal{L}_\xi \gamma(\delta) \big].
    \label{differences}
\end{equation}
Hence, following Ref.~\onlinecite{Iyer}, suitable fall-off conditions must be imposed on the dynamical fields so that the boundary contribution in Eq.~\eqref{differences} vanish at spatial infinity. These conditions are specified below. Therefore, the relevant information about the conserved charges at infinity can be obtained by working with $ \Omega (\delta, \mathcal{L}_\xi) $, which is the central object used in what follows. The evaluation of the charges at infinity is addressed in the next subsection.

\subsection{Decay rates}\label{CInf}

The region of interest is spatial infinity, so it is important to determine the fall-off rates of all dynamical fields. In the metric formalism in $n$ dimensions, $n>3$, the fall-off rates at spatial infinity for asymptotically flat spacetimes are given by~\cite{10.1063/1.3166141,Hollands2004Asymptotic}
\begin{equation}
     g_{ab} = \eta_{ab} + O(1/r^{n-3}),
    \label{falloffrates}
\end{equation}
where $r$ is a radial coordinate such that $\text{i}^0$ is associated with the limit $r\to \infty$. Using the prescription given in Refs.~\onlinecite{Corichi1,AshtekarEngleSloan2008}, the veilbein must satisfy
\begin{equation}
    e^\mu = {}^{(0)}e^\mu + \frac{{}^{(1)}e^\mu}{r^{n-3}} + O(1/r^{n-2}),
    \label{nbeinfalloff}
\end{equation}
where the superindex on the left denotes the order of the expansion. It is worth noting that this expansion leads to the correct expansion of the metric as $g_{ab} = \eta_{\mu \nu} e^{\mu}_a e^{\nu}_b$ and $\eta_{\mu \nu}$ fixes the zero order expansion of the veilbein $(\eta_{ab} = \eta_{\mu \nu} {}^{(0)}e^\mu_a {}^{(0)}e^\nu_b)$. In addition, the spin connection is assumed to have the following decay rate, thus
\begin{equation}
    \omega^{\mu \nu} = {}^{(0)} \omega^{\mu \nu} + \frac{{}^{(1)}\omega^{\mu \nu}}{r^{n-3}} + \frac{{}^{(2)}\omega^{\mu \nu}}{r^{n-2}} + O(1/r^{n-1}).
    \label{SCfalloff}
\end{equation}
Additionally, torsion must vanish to zero-th order. Given $T^\mu = \D e^\mu$, the zero-th order equation is~\cite{Corichi1}
\begin{equation}
    \de \left[ {}^{(0)}e^\mu \right]  + {}^{(0)} \omega^{\mu}_{\ \nu} \wedge {}^{(0)}e^\nu = 0.
    \label{torsionzeroorder}
\end{equation}
It is clear that $\mathrm{d}\left[{}^{(0)}e^\mu \right]= 0$, and therefore ${}^{(0)}\omega^{\mu\nu} = 0$. Thus, to obtain well-defined charges at spatial infinity (finite and gauge independent), ${}^{(1)}\omega^{\mu\nu}$ must vanish. 

Furthermore, since spacetime becomes flat at spatial infinity, the spin connection at leading order must be determined solely by the veilbeins. Consequently, from the torsion equation at order $1/r^{n-3}$ it follows that ${}^{(1)}\omega^{\mu\nu}$ cannot be fixed by the veilbeins, and therefore ${}^{(1)}\omega^{\mu\nu}$ corresponds to a pure gauge contribution. In particular, it must satisfy the following relation~\cite{Corichi1,AshtekarEngleSloan2008}:
\begin{equation}
    \mathrm{d}\!\left[\frac{{}^{(1)}e^\mu}{r^{n-3}} \right] 
    = -\frac{1}{r^{n-2}}\,{}^{(2)}\omega^\mu_{\ \nu} \wedge {}^{(0)}e^\nu.
    \label{conn2order}
\end{equation}
In addition, Eq.~\eqref{conn2order} implies that torsion decays as $T^\mu = O(1/r^{n-1})$.

When including matter fields, the conserved charges at spatial infinity can be computed using
\begin{equation}
    \delta H [\xi] = \int_{\text{i}^0} \ \delta Q(\xi) - \text{i}_\xi \theta (\delta),
    \label{chargesatinfinity}
\end{equation}
where one can use either the standard Noether charges and the boundary term, or their gauge invariant versions, as the asymptotic fall-off rates of the metric are sufficiently rapid to suppress the contribution from \(\Delta \Omega (\delta, \mathcal{L}_\xi)\).

Hence, the variation of the Hamiltonian for the gravitational sector can be written as
\begin{widetext}
    \begin{eqnarray}
        \delta H[\xi] &=& \int_{i^{0}} \ \delta \omega^{\mu \nu} \wedge \text{i}_\xi H_{\mu \nu}^{(0)} 
        + \text{i}_\xi\omega^{\mu \nu} \wedge \delta H_{\mu \nu}^{(0)} 
        + \delta e^{\mu} \wedge \text{i}_\xi E_{\mu }^{(0)} 
        + \xi^{\mu}  \delta E_{\mu }^{(0)} \nonumber \\
        &&+ \text{i}_\xi \star R^{\mu \nu} \wedge \delta H_{\mu \nu}^{(1)} 
        - (-1)^{n} \delta \star R^{\mu \nu} \wedge \text{i}_\xi H_{\mu \nu}^{(1)}
        + \text{i}_\xi \star T^{\mu } \wedge \delta E_{\mu }^{(1)} 
        - (-1)^{n} \delta \star T^{\mu} \wedge \text{i}_\xi E_{\mu}^{(1)} \nonumber \\
        && +  \sum_{j=1}^{N-1} \left[ \text{i}_\xi \star \D^{j} \star R^{\mu \nu} \wedge \delta H^{(j+1)}_{\mu \nu} 
        - (-1)^{(n-1)j}\delta \star \D^{j} \star R^{\mu \nu} \wedge \text{i}_\xi H^{(j+1)}_{\mu \nu}\right] \nonumber \\
        && +  \sum_{j=1}^{M-1} \left[ \text{i}_\xi \star \D^{j} \star T^{\mu} \wedge \delta E^{(j+1)}_{\mu } 
        - (-1)^{(n-1)j}\delta \star \D^{j} \star T^{\mu } \wedge \text{i}_\xi E^{(j+1)}_{\mu }\right]. 
        \label{Hamiltonianvariation}
    \end{eqnarray}  
\end{widetext}
The asymptotic behavior of the dynamical fields determines which terms in Eq.~\eqref{Hamiltonianvariation} contribute at spatial infinity (take into the account that the integration measure introduces a factor \(r^{n-2}\) when integrating over a topological sphere at spatial infinity). The finiteness of the Hamiltonian variation requires that \(H_{\mu \nu}^{(I)}\) and \(E_{\mu}^{(I)}\) exhibit suitable fall-off conditions. Under these assumptions, only the first four terms in Eq.~\eqref{Hamiltonianvariation} decay slowly enough to contribute to the conserved quantities. Consequently, to ensure a finite result, the following asymptotic behavior is assumed:
\begin{eqnarray}
    H_{\mu \nu}^{(0)} &=& \bar{H}_{\mu \nu}^{(0)} + O(1/r^{n-3}),  \\
    E_{\mu}^{(0)} &=& \frac{\bar{E}_{\mu }^{(0)}}{r^{n-2}} + O(1/r^{n-1}).
    \label{Hdecay}
\end{eqnarray}
Now, from Eqs.~\eqref{IGTTH0} and \eqref{EGTNH0}, one can see that the theory admits asymptotically flat solutions if the decay has the form given in Eq.~\eqref{Hdecay}. For example, in four dimensions, the terms that are related with torsion decay sufficiently rapid and do not contribute in the expression of the charges. In general, the charge at infinity with the previous assumptions is
\begin{equation}
    \delta H[\xi] = \int_{\text{i}^0} \ \delta \omega^{\mu \nu} \wedge \text{i}_\xi H_{\mu \nu}^{(0)} 
        + \xi^{\mu}  \delta E_{\mu }^{(0)}.
    \label{ChargesatInf}
\end{equation}

The charges at infinity can be deduced from Eq.~\eqref{ChargesatInf} and depend on $\xi$. For a stationary black hole with a bifurcation surface, the relevant Killing vector \( \xi \) can be expressed as~\cite{wald}
\begin{equation}
\xi = t + \Omega_H \phi,
\label{Temporal}
\end{equation} 
where \( t \) is the stationary Killing vector field, \( \phi\) is the rotational Killing vector field, and $\Omega_H$ is the angular velocity of the horizon~\cite{WaldB}. Thus,
\begin{equation}
    \int_{\text{i}^0} \ \delta \omega^{\mu \nu} \wedge \text{i}_\xi H_{\mu \nu}^{(0)} 
        + \xi^{\mu}  \delta E_{\mu }^{(0)} = \delta M - \Omega_H \delta J,
    \label{massangularmomentum}
\end{equation}
where $\delta M = \delta H[t]$ and $\delta J =- \delta H[\phi]$.

Matter fields can be included in a straightforward manner using Eq.~\eqref{Noetherchargemat} and the boundary term given in Eq.~\eqref{BTMatter}. Luckily, the contribution of matter has the same structure than its gravitational counterpart. Thus,
\begin{widetext}
    \begin{eqnarray}
        \int_{\text{i}^0}\delta Q_{M} - \text{i}_\xi \theta_M (\delta) &=& \int_{\text{i}^0} \delta \omega^{\mu \nu} \wedge \text{i}_\xi \mathcal{H}_{\mu \nu}^{(0)} + \xi^{\mu} \delta \mathcal{E}_\mu^{(0)} +\text{i}_\xi \Phi^{I} \wedge \delta E_{I}^{(1)} 
        - (-1)^p \delta\Phi^{I} \wedge \text{i}_\xi E_{I}^{(1)} \nonumber \\ 
        &&+\text{i}_\xi \star \Phi^{I} \wedge \delta G_{I}^{(1)} - (-1)^{n-p}\delta \star \Phi^{I} \wedge \text{i}_\xi G_{I}^{(1)}  \nonumber \\
        &&+ \sum_{j=1}^{k-1} \bigg{[}\text{i}_{\xi} \star \D^{j} \Phi^{I} \wedge \delta E_{I}^{(j+1)} - (-1)^{\alpha(p,j-1)}\delta \star \D^{j} \Phi^{I} \wedge \text{i}_\xi E_{I}^{(j+1)} \bigg{]} \nonumber \\
        &&+ \sum_{j=1}^{l-1}\bigg{[} \text{i}_{\xi} \star \D^{j} \star \Phi^{I} \wedge \delta H_{I}^{(j+1)} - (-1)^{\alpha(n-p,j-1)} \delta \star \D^{j} \star \Phi^{I} \wedge \text{i}_\xi H_{I}^{(j+1)}\bigg{]}\nonumber \\ 
        && + \text{i}_\xi \mathit{A}^{I}_{\ J} \wedge \delta [H^{(0)}]_{I}^{\ J}  + \delta \mathit{A}^{I}_{\ J} \wedge \text{i}_\xi [H^{(0)}]_{I}^{\ J} \nonumber \\
        && + \text{i}_\xi \star \mathit{F}^{I}_{\ J} \wedge \delta [H^{(1)}]_{I}^{\ J} - (-1)^{n}   \delta \star \mathit{F}^{I}_{\ J} \wedge \text{i}_\xi [H^{(1)}]_{I}^{\ J} \nonumber \\
        && + \sum_{j=1}^{h-1} \bigg{[} \text{i}_{\xi} \star \D^{j} \star \mathit{F}^{I}_{\ J} \wedge \delta  [H^{(j+1)}]_{I}^{\ J} - (-1)^{j(n-1)}  \delta \star \D^{j} \star \mathit{F}^{I}_{\ J} \wedge \text{i}_\xi  [H^{(j+1)}]_{I}^{\ J}\bigg{]}.
        \label{chargesmatter}
    \end{eqnarray}
\end{widetext}
As occurs in the gravity theory, all contributions associated with higher-derivative curvature and torsion terms can be discarded, leaving only the above displayed terms as potential contributions to the asymptotic charges. The remaining task is therefore to determine which of these terms survive once appropriate asymptotic conditions are imposed on the matter fields.

The matter fields are required to decay sufficiently rapidly so that the corresponding charges remain finite. In analogy with the gravitational case, all terms involving derivatives of the matter fields must vanish in the asymptotic region. Under these conditions, the dominant contributions from Eq.~\eqref{chargesmatter} are
\begin{widetext}
\begin{eqnarray}
    \int_{\text{i}^0}\delta Q_{M} - \text{i}_\xi \theta_M (\delta)   &=& \int_{\text{i}^0} \text{i}_\xi \Phi^{I} \wedge \delta E_{I}^{(1)} 
        - (-1)^p \delta\Phi^{I} \wedge \text{i}_\xi E_{I}^{(1)} +\text{i}_\xi \star \Phi^{I} \wedge \delta G_{I}^{(1)} - (-1)^{n-p}\delta \star \Phi^{I} \wedge \text{i}_\xi G_{I}^{(1)}  \nonumber \\
        && + \text{i}_\xi \mathit{A}^{I}_{\ J}  \delta [H^{(0)}]_{I}^{\ J}  + \delta \mathit{A}^{I}_{\ J} \wedge \text{i}_\xi [H^{(0)}]_{I}^{\ J}+ \delta \omega^{\mu \nu} \wedge \text{i}_\xi \mathcal{H}_{\mu \nu}^{(0)} + \xi^{\mu} \delta \mathcal{E}_\mu^{(0)} .
    \label{contributionmatter}
\end{eqnarray}
\end{widetext}

At this stage, a specific matter model must be introduced. Nevertheless, if the fields exhibit the asymptotic behavior~\cite{PhysRevD.99.084007}
\begin{eqnarray}
    \Phi^{I} &=& \frac{\bar{\Phi}^{I} }{r^{n-3}} + O(1/r^{n-2}), 
    \label{Falloffmatter1} \\
    \mathcal{A}^{I}_{J} &=& \frac{\bar{\mathcal{A}}^{I}_{\ J} }{r^{n-3}} + O(1/r^{n-2}),
    \label{falloffmatter2}
\end{eqnarray}
then, the matter contribution to the Hamiltonian is finite. Clearly, if the dynamical fields exhibit a faster fall-off at infinity, its contribution to the asymptotic charges vanishes. Therefore, the decay assumed here represents the minimal fall-off rate that can contribute to conserved charges.

The functionals appearing in Eq.~\eqref{contributionmatter} must satisfy
\begin{eqnarray}
    E_{I}^{(1)}  &=& \frac{\bar{E}^{(1)}}{r^{n-2}}+ O(1/r^{n-3}), 
    \label{Falloffmatter3} \\
    G_{I}^{(1)}  &=& \frac{\bar{G}^{(1)}}{r^{n-2}}+ O(1/r^{n-3}), \\
    \label{falloffmatter4}
    [H^{(0)}]_{I}^{\ J}  &=& \frac{[\bar{H}^{(0)}]_I^{\ J}}{r^{n-2}}+ O(1/r^{n-3}).
\end{eqnarray}
Now, as it is discussed in Appendix \ref{EOMGTG}, the previous objects are determined by the partial derivatives of the Lagrangian with respect to the matter fields and their covariant derivatives. Thus,
\begin{equation}
    \int_{\text{i}^0} \delta Q_{M} - \text{i}_\xi \theta_{M}(\delta)
    = V^{I}{}_{J} \, \delta \mathcal{Q}_{I}{}^{J}+ \delta \omega^{\mu \nu} \wedge \text{i}_\xi \mathcal{H}_{\mu \nu}^{(0)} + \xi^{\mu} \delta \mathcal{E}_\mu^{(0)} ,
    \label{MatterfieldCharges}
\end{equation}
where $\mathcal{Q}_{I}{}^{J} = \int_{\text{i}^0} [H^{(0)}]_{I}{}^{J}$ is the generalization of the electric charge in a generic gauge theory. 
Consequently, under the assumed fall-off conditions, the only contribution from matter fields to the conserved charges at spatial infinity arises from the gauge sector.

In fact, the analogue of the electrostatic potential of Maxwell's theory is given by $\text{i}_\xi A^{I}{}_{J}$, which, depending on the underlying theory, may be pure gauge~\cite{DanielWald}. Note that
\begin{equation}
   \text{i}_\xi A^{I}_{J} = V^I_{\ J} + O(1/r^{n-3}),
  \label{nodynamical}
\end{equation}
where $V^I_{\ J}$ is the ``electrostatic potential" at spatial infinity, which is constant.

Moreovoer, the second and third terms in the right-hand side of Eq.~\eqref{MatterfieldCharges} encode possible matter contributions to the gravitational charges.

In the present framework, a minimal coupling is understood as one in which the spin connection appears in the matter Lagrangian exclusively through the covariant derivatives of the matter fields. In contrast, a nonminimal coupling arises whenever the matter Lagrangian depends explicitly on the curvature and/or torsion. Then, note that for minimally coupled matter fields, $\mathcal{H}_{\mu\nu}^{(0)}$ vanishes and these contributions are typically absent. By contrast, nonminimal couplings to the gravitational sector may generate nontrivial $\mathcal{H}_{\mu\nu}^{(0)}$ and $\mathcal{E}_{\mu}^{(0)}$, leading to corrections to the Hamiltonian energy and angular momentum.

At this stage, the general expressions required to derive the first law of black hole thermodynamics in a broad class of gravitational theories coupled to matter have been found. An example is presented in the next section.

\section{Example}

The formalism developed in the previous sections is illustrated with a simple example. Consider four-dimensional Einstein--Cartan gravity coupled nonminimally to a scalar field through a Gauss--Bonnet term. In four spacetime dimensions, the Gauss--Bonnet term is topological. However, the nonminimal coupling produces nontrivial modifications. Concretely, the action is given by
\begin{eqnarray}
    S[e,\omega; \phi] &=& \frac{1}{2\kappa} \int \epsilon_{\mu \nu \alpha \beta} \, R^{\mu \nu} \wedge e^{\alpha} \wedge e^{\beta} \nonumber \\
    && + \frac{\alpha}{4} \int \ \phi^{2} \epsilon_{\mu \nu \rho \sigma} R^{\mu \nu} \wedge R^{\rho \sigma} \nonumber \\
    &&- \frac{1}{2} \int \ \de \phi \wedge \star \de \phi, 
    \label{ActionECSGB}
\end{eqnarray}
where $\kappa = 16\pi G$, $\alpha$ is a coupling constant, and $\phi$ is a scalar field. Since the scalar representation of the Lorentz group is trivial, one has $(\mathfrak J^{\mu\nu})^{I}{}_{J}=0$ and therefore $\D\phi=\de\phi$.

The corresponding equations of motion follow directly from Eqs.~\eqref{EomemGV}, \eqref{hodeomA4}, \eqref{cuantities}, \eqref{Spindensit} and \eqref{EOMMAtterPhi}:
\begin{eqnarray}
     \frac{1}{\kappa} \epsilon_{\mu \nu \rho \sigma} R^{\nu \rho} \wedge e^\sigma &=&- \frac{1}{2}\text{i}_\mu \star \de \phi \wedge \de \phi - \frac{1}{2} \text{i}_\mu\de \phi \wedge  \star \de \phi, \nonumber \\
     &&    \label{ECSGBeom1}\\
     \frac{1}{\kappa} \epsilon_{\mu \nu \rho \sigma} T^{\rho} \wedge e^\sigma  &=&- \alpha (\phi\de\phi) \epsilon_{\mu \nu \rho \sigma} R^{\rho \sigma}\,    \label{ECSGBeom2}
 \\
     \de \star \de \phi &=& - \frac{\alpha}{2} \phi \epsilon_{\mu \nu \rho \sigma } R^{\mu \nu} \wedge R^{\rho \sigma}.
    \label{ECSGBeom3}
\end{eqnarray}

The entropy is obtained from the horizon contribution using Eq.~\eqref{Bifurcationpartxi}. For the present theory one finds
\begin{equation}
     - \int_{\mathcal{B}} \mathring{\nabla}^\mu \xi^\nu \left[  \frac{1}{2 \kappa}   \epsilon_{\mu \nu \rho \sigma} e^\rho \wedge e^\sigma + \frac{\alpha \phi^2}{2 }  \epsilon_{\mu \nu \rho \sigma}R^{\rho \sigma}  \right].
    \label{EntroECSGBgrav}
\end{equation}
The asymptotic charges follow from Eqs.~\eqref{ChargesatInf} and \eqref{MatterfieldCharges}. Under the asymptotic conditions assumed throughout this work, the scalar field does not contribute explicitly at spatial infinity, and the only surviving term is
\begin{equation}
    \int_{\text{i}^0} \ \frac{1}{\kappa}\delta \omega^{\mu \nu} \wedge \epsilon_{\mu \nu \rho \sigma} \xi^\rho e^\sigma.
    \label{iocharges}
\end{equation}
Thus the first law of black hole thermodynamics is
\begin{widetext}
\begin{eqnarray}
    \chi \delta \int_{\mathcal{B}}n^{\mu \nu} \left[  \frac{1}{2 \kappa}   \epsilon_{\mu \nu \rho \sigma} e^\rho \wedge e^\sigma + \frac{\alpha \phi^2}{2 }  \epsilon_{\mu \nu \rho \sigma}R^{\rho \sigma}  \right] = \int_{\text{i}^0} \ \frac{1}{\kappa}\delta \omega^{\mu \nu} \wedge \epsilon_{\mu \nu \rho \sigma} \xi^\rho e^\sigma.
    \label{BHMECSGBgrav}
\end{eqnarray}
\end{widetext}

Note that the asymptotic charges remain unchanged with respect to Einstein--Cartan gravity \cite{Corichi1}. The only modification when compared with this theory appears in the entropy, and it is due to the nonminimal coupling between the scalar field and the Gauss--Bonnet term, as shown by the fact that it vanishes when $\alpha=0$.

\section{Conclusions}\label{sec:conclusions}

The formulation of the first law of black hole thermodynamics within the formalism where gravity and matter fields are described with differential forms has been revisited. A generalization of the Iyer–Wald procedure has been introduced that is manifestly covariant, gauge independent, and formulated entirely in terms of differential forms defined on spacetime. As such, the derivation avoids steps associated with lifting and projecting the fields into the principal fiber bundle, which is the only known alternative that retains gauge invariance.

The construction presented here leads to a gauge-invariant presymplectic current which explains a correction previously introduced for a particular theory in Ref.~\onlinecite{Simone}. Beyond its role in black hole thermodynamics, this structure may also prove useful in contexts where differential form are employed in the quantization of fields on curved backgrounds.

The gauge-invariant formalism developed here is applied to a broad class of gravity and matter theories formulated in terms of differential forms. The main differences with respect to the metric formulation arise in the definition of gauge-invariant charges and in the structure of the boundary term. These quantities encode the information relevant for the black hole entropy and the conserved charges at spatial infinity.

A general expressions for the quantities entering the first law is obtained for a generic Lagrangian. In particular, when gauge invariance is imposed in the gravity sector, the Lagrangian can depend only on the vielbein, the curvature 2-form, the torsion 2-form, and their covariant derivatives (no nondynamical fields are considered). This formulation offers conceptual and technical advantages over the metric formalism; most notably, torsion is incorporated naturally, and many computations become more transparent. Matter fields can also be included straightforwardly contributing in a clear manner to both, the entropy and the charges at infinity.

Another significant outcome of this work is that the generic structure of the first law of black hole thermodynamics depends crucially on whether the coupling between gravity and matter is minimal or nonminimal. In particular, $\mathcal{H}_{\mu\nu}^{(0)}$ vanishes for minimally coupled theories but is generally nonzero for nonminimally coupled ones, as illustrated by the example.

Whether a given theory admits solutions satisfying the assumed conditions, namely, stationarity, the existence of a bifurcate Killing horizon, and asymptotic flatness, is not verified here. This must be checked on a case-by-case basis. Indeed, a given theory may admit several distinct families of solutions satisfying these conditions, making the uniqueness established here particularly interesting.

In addition, interpretative aspects have been clarified, including the relation between the Hamiltonian and the presymplectic current in the used framework. In particular, the presymplectic current determines the variation of the Hamiltonian and enforces the Hamilton equations of motion; therefore, it encodes the variation of the conserved charges at infinity. The main difference with respect to the metric formalism lies in the choice of presymplectic current as in the present case it is necessary to use the gauge-independent version $\hat{\Omega}$. This current differs from the one in the metric formalism by a boundary term that vanishes at spatial infinity and thus does not affect the charges defined there. However, the use of $\hat{\Omega}$ becomes essential when evaluating quantities on the bifurcation surface, which remains invariant under ambiguities associated with the boundary term. Moreover, the ambiguity arising from the Lagrangian is controlled by the presymplectic current, and the remaining ambiguity in the charge can be fixed by requiring consistency when the vector field involved is multiplied by a function. Consequently, the geometric derivation of the first law of black hole thermodynamics presented here is invariant under all these ambiguities.


The treatment at spatial infinity requires appropriate fall-off conditions of the fields. In this work, a Coulomb-like decay is assumed for the relevant matter fields, which allows one to obtain explicit expressions for the charges at infinity. However, if the theory admits solutions with different asymptotic behavior, the resulting charges could be modified. Under the Coulomb-like fall-off conditions for the gravitational and matter sectors, the Hamiltonian yields general expressions for the energy and angular momentum. In such cases, torsion contributes to the charges only when the decay of $E_{\mu}^{(0)}$, defined in Eq.~\eqref{EGTNH0}, is sufficiently slow. Thus, the question of how torsion affects the energy reduces to constructing a Lagrangian for which $E_{\mu}^{(0)}$ exhibits the desired asymptotic behavior. For the matter sector, in turn, the only contributions to the first law arise from gauge fields, thereby generalizing the standard electrostatic potential term in Maxwell theory.

An important open question concerns the connection between the temperature appearing in the first law of black hole thermodynamics and the temperature associated with Hawking radiation. While the geometric derivation presented here is entirely classical, Hawking's result relies on quantum field theory in curved spacetime. A natural direction for future work is to compare the temperature obtained from the present formalism with the Hawking temperature for known black hole solutions in modified gravitational theories and determine whether the agreement found in general relativity persists beyond Einstein gravity. Another interesting connection to explore (suggested by Johas Morales) would be between the modification introduced here to render the presymplectic current gauge invariant and the correction introduced by Dirac to the Poisson bracket to address a similar problem.

Overall, this work shows that the first law of black hole thermodynamics is fundamentally geometric and applies to an enormous class of theories. In addition, it illustrates the advantages of carrying out these calculations using differential forms.

\appendix
\section{Gauge parameter}\label{GaugeParam}

This appendix is devoted to the construction of a particular field-dependent gauge parameter relevant to black hole thermodynamics. Let $\mathring{\nabla}_\mu$ be such that $ \mathring{\nabla}_\nu \xi^\mu = e^{a}_\nu e^{\mu}_b \mathring{\nabla}_a \xi^b$, where $\mathring{\nabla}_a$ is the torsion-free covariant derivative. When acting on $\xi^\mu = e^\mu(\xi)$, which is an algebra-valued $0$-form, it produces
\begin{equation}\label{nablaGreek}
\mathring{\nabla}_\nu \xi^\mu = \text{i}_\nu\left( \de \xi^\mu + \mathring{\omega}^{\mu}_{\ \rho} \xi^\rho\right).
\end{equation}
Here, $\text{i}_\nu$ is the interior derivative along the dual vector associated with $e^\nu$. What is more, a direct application of Cartan's magic formula yields
\begin{equation}
 \mathcal{L}_{\xi} e^{\mu} = \de \xi^{\mu} - \text{i}_\xi \mathring{\omega}^{\mu}_{\ \nu} \wedge e^\nu + \mathring{\omega}^{\mu}_{\ \nu} \xi^\nu= \left(\mathring{\nabla}_\nu \xi^{\mu} - \text{i}_\xi \mathring{\omega}^{\mu}_{\ \nu} \right) e^\nu,
 \label{Killingtetrad}
\end{equation}
with $\mathring{\omega}^{\mu}_{\ \nu}$ the torsion-free part of $\omega^{\mu}_{\ \nu}$. Then, one can readily show that
\begin{equation}
e^{[\mu}_a \mathcal{L}_\xi e^{\nu]a} = \mathring{\nabla}^{[\mu} \xi^{\nu]} + \text{i}_\xi \mathring{\omega}^{\mu \nu} =: \lambda_\xi^{\mu \nu},
\label{GaugeTKilling}
\end{equation}
where the brackets denote antisymmetrization of indices. Notice that $\lambda_\xi^{\mu \nu}=\lambda_\xi^{[\mu \nu]}$ can be used as a gauge parameter.

\section{Lagrangian invariance} \label{InvLag}

Throughout this work, it is assumed that the Lagrangian contains no exact terms, which do not affect the equations of motion. This choice provides a consistent prescription for applying the formulas that are developed. Moreover, as shown in Appendix~\ref{ChargesAtInf}, the addition of an exact term to the Lagrangian does not modify the presymplectic current or the conserved charges.

In this appendix it is shown that the presence of an exact ther does not affect the form of the first law of black hole thermodynamics. Clearly, two Lagrangians that differ by an exact term yield different boundary terms. In fact, $\theta$ would differ by an exact form $\de\beta$, and thus $\Omega$ differs by a boundary term. However, as discussed around Eq.~\eqref{differences}, this difference produces no effects for fields that decay fast enough at spatial infinity.

The Holst Lagrangian provides an explicit illustration of this case. The Lagrangian is given by
\begin{equation}
   R^{\mu\nu}\wedge e_\mu\wedge e_\nu = T^\mu\wedge T_\mu - \de\big(e^\mu\wedge T_\mu\big). 
\end{equation}
Writing the Holst term using curvature or, alternatively, using $T^\mu \wedge T_\mu$ plus the exact form does not alter the bulk equations, but it changes the boundary term. The difference in the two treatments is given by $-\de(e^\mu\wedge \delta e_\mu)$ in the general action variation, which produces a boundary contribution to \(\Delta\Omega\) that has the form of Eq.~\eqref{differences}. Luckily, for the fall-off conditions adopted in this work, the resulting boundary contribution decays sufficiently rapid when approaching spatial infinity, and therefore does not modify the charges.

In general it follows that any Lagrangian yielding the same equations of motion, leads to identical charges at infinity for the assumed decays. Furthermore, the invariance of the first law of black hole thermodynamics, particularly on the bifurcation surface, is guaranteed, as can be seen from the form of $\hat{\Omega}(\delta, \mathcal{L}_\xi)$.

In conclusion, Lagrangians with the same dynamical content give rise to the same first law of black hole thermodynamics. Moreover, once the Noether charge is fixed as in Eq.~\eqref{interiorL0}, the remaining ambiguities discussed in Ref.~\onlinecite{Iyer} do not modify the form of the first law.

\section{Matter equations of motion} \label{EOMGTG}

In this appendix, the expressions for the energy--momentum and spin density appearing in the gravitational field equations, together with the equations of motion of the matter fields are derived.

The energy-momentum $(n-1)$-form is deduced using a recursive method analogous that that of Sec.~\ref{SecGTGl}. The method requires starting from the term with the highest order of derivatives, $k$. Let
\begin{equation}
    U_{I}^{(k)} = \frac{\partial \mathcal{L}_{M}}{ \partial \star \D^{k}  \Phi^{I}}.
    \label{Materia}
\end{equation}
The variation associated with it satisfies
\begin{eqnarray}
 \delta \star \D^{k} \Phi^{I}  \wedge U_{I}^{(k)} &=& (-1)^{u(p,k)} \delta \D^k \Phi^{I}  \wedge \star U_I^{(k)} \nonumber \\
 && - \star\left(\delta e^\mu \wedge \text{i}_\mu \D^{k} \Phi^{I} \right) \wedge U_I^{(k)} \nonumber \\
 &&+ \delta e^\mu \wedge \text{i}_\mu \star \D^k \Phi^I \wedge U_I^{(k)}
 \label{VarofDPhi}
\end{eqnarray}
where $u(p,k)=k(n-1)+p(n-p)$, which arises from manipulating the Hodge star. The first term in Eq.~\eqref{VarofDPhi} contributes to the variation of $\D \Phi^I$. The other two terms contribute to the energy momentum $(n-1)$-form.

Defining
\begin{eqnarray}
    E_{I}^{(k)} &=& \frac{\partial \mathcal{L}_{M}}{ \partial \D^{k}\Phi^{I}} +  (-1)^{u(p,k)} \star U_I^{(k)},
    \label{MRecursive}
\end{eqnarray}
the variation of $\D \Phi^I$ takes the form
\begin{eqnarray}
    \delta \D^{k} \Phi^I \wedge E_I^{(k)} &=& \de \left[\delta \star \D^{k-1} \Phi^I \wedge E_I^{(k)} \right] \nonumber \\
    &&+ (-1)^{v (p,k)} \delta \star \D^{k-1} \Phi^I \wedge \D E_I^{(k)} \nonumber \\
    &&+ \delta \mathit{A}^I_J \star \D^{k-1} \Phi^J \wedge E_I^{(k)} \nonumber \\
    && + \delta \omega^{\mu \nu} \wedge \left[ \frac{1}{2}(\mathfrak{J}_{\mu\nu})^{I}_{\ J} \star \D^{k-1} \Phi^J \wedge E_I^{(k)} \right], \nonumber \\
    \label{varEI}
\end{eqnarray}
where $v(p,k)=(k+1)(n-1)-p$. The first term in Eq.~\eqref{varEI} leads to a boundary term that appears in Eq.~\eqref{BTMatter}, the second contributs to the variation of one lower order, and the last two terms appear in the equations of motion of the gauge fields.

It is then possible to compute the variation of $\delta \Phi^I$ by iterating recursively and defining 
\begin{eqnarray}
    U^{(k-1)}_I &=&  \frac{\partial \mathcal{L}_{M}}{ \partial \star \D^{k-1}  \Phi^{I}} + (-1)^{v (p,k)}\D E_I^{(k)},
    \label{Uk-1}
\\
    E^{(k-1)}_I &=&  \frac{\partial \mathcal{L}_{M}}{ \partial  \D^{k-1}  \Phi^{I}} + (-1)^{u (p,k-1)}\star U_I^{(k-1)}.
    \label{Ek-1}
\end{eqnarray}

The variations of the terms proportional to $\star \Phi^{I}$ and $\star \mathit{F}^{I}_{\ J}$ is straightforward. These variations correspond to $(n-p)$- and $(n-2)$-forms, respectively. Accordingly, for the highest order, the following quantities are introduced:
\begin{eqnarray}
    G_{I}^{(l)} &=& \frac{\partial \mathcal{L}_{M}}{\partial \star \D^{l} \star \Phi^{I}}, 
    \label{MRecursivestar}
\\
    {[I^{(h)}]}_I^{\ J} &=& \frac{\partial \mathcal{L}_{M}}{\partial \star \D^{h} \star F^{I}_{\ J}}.
    \label{MRecursivestar1}
\end{eqnarray}
Thus, 
\begin{equation}
    P_{I}^{(l)} = \frac{\partial \mathcal{L}_M}{\partial  \D^{l} \star \Phi^I} + (-1)^{v (n-p,l)} \star G_I^{(l)},
    \label{AnalogoMatterstar} 
\end{equation}
and
\begin{equation}
    [H^{(h)}]_I^{ \ J} = \frac{\partial \mathcal{L}_M}{\partial  \D^h \star F^I_{\ J}} + (-1)^{h(n+1)} \star  [I^{(h)}]_I^{\ J}.
    \label{CurvatureAnalog} 
\end{equation}
The following order is then given by
\begin{eqnarray}
    G_{I}^{(l-1)} = \frac{\partial \mathcal{L}_M}{\partial \star \D^{l-1} \star \Phi^I} + (-1)^{v (n-p,l)} \D P_I^{(l)},
    \label{starmatter}
\\
    P_{I}^{(l-1)} = \frac{\partial \mathcal{L}_M}{\partial  \D^{l-1} \star \Phi^I} + (-1)^{u (n-p,l-1)} \star G_I^{(l-1)}.
    \label{starmatter3}
\end{eqnarray}
This process is continued until the lowest order is attained, which yields
\begin{equation}
    G_I^{(0)} = \frac{\partial \mathcal{L}_{M}}{ \partial \star \Phi^{I}} + (-1)^{v (n-p,0)}\D P_I^{(1)}.
    \label{0order2}
\end{equation}

The variation associated with $\star \mathit{F}^{I}_{\ J}$ is completely analogous to the variation of the Hodge dual of the curvature $\star R^{\mu \nu}$ in the gravitational sector. It is driven by recursive expression involving
\begin{eqnarray}
     [I^{(h-1)}]_{I}^{\ J}  &=&\frac{\partial \mathcal{L}_{M}}{\partial \star \D^{l-1} \star F_{I}^{\ J}} + (-1)^{h(n+1)} \D[H^{(h)}]_{I}^{\ J} , \nonumber \\
     && \\
    \label{H1Matter} 
    [H^{(h-1)}]_{I}^{\ J} &=& \frac{\partial \mathcal{L}_M}{\partial \D^{h-1} \star F^{I}_{\ J}} \nonumber \\ 
    &&+ (-1)^{(h-1)(n+1)}\star [I^{(h-1)}]_{I}^{\ J} . 
    \label{H1Matter3} 
\end{eqnarray}

With all these expressions, it is straightforward to construct the energy-momentum tensor and the spin density, which are respectively given by
\begin{widetext}
\begin{eqnarray}
-\tau_\mu &=& \Xi_\mu + \sum^{k}_{j=1}  \bigg{[}\text{i}_\mu  \star \D^{j} \Phi^{I} \wedge U^{(j)}_{I}-(-1)^{u(p,j)} \text{i}_\mu   \D^{j} \Phi^{I}\wedge \star U^{(j)}_{I} \bigg{]} + \text{i}_\mu \star \Phi^{I} \wedge G^{(0)}_{I} - (-1)^{u(p,0)} \text{i}_\mu \Phi^{I} \wedge \star  G^{(0)}_{I}\nonumber \\
&& + \sum^{l}_{j=1} \bigg{[} \text{i}_\mu  \star \D^{j} \star \Phi^{I} \wedge G^{(j)}_{I}-(-1)^{u(p,j)} \text{i}_\mu   \D^{j} \star \Phi^{I}\wedge \star G^{(j)}_{I} \bigg{]} 
+ \text{i}_\mu   \star F^{I}_{\ J} \wedge [H^{(0)}]_{I}^{\ J}- \text{i}_\mu  F^{I}_{\ J}\wedge \star [H^{(0)}]_{I}^{\ J}  \nonumber \\ 
&& + \sum^{h}_{j=1} \bigg{[} \text{i}_\mu  \star \D^{j} \star F^{I}_{\ J} \wedge [H^{(j)}]_{I}^{\ J}-(-1)^{jn} \text{i}_\mu   \D^{j} \star F^{I}_{\ J}\wedge \star [H^{(j)}]_{I}^{\ J} \bigg{]} ,
 \label{cuantities}
\end{eqnarray}
\begin{equation}
    -S_{\mu \nu} = \Xi_{\mu \nu} + \frac{1}{2} (\mathfrak{J}_{\mu \nu})^{I}_{\ J} \left[ \star \Phi^{J} \wedge P_{I}^{(1)} + \Phi^{J} \wedge E_{I}^{(1)} \right]+  \frac{1}{2} (\mathfrak{J}_{\mu \nu})^{I}_{\ J} \left[ \sum_{i=1}^{k-1} \star \D^{i} \Phi^{J} \wedge E_{I}^{(i+1)}  
 + \sum_{i=1}^{l-1} \star \D^{i} \star \Phi^{J} \wedge P_{I}^{(i+1)}\right],
    \label{Spindensit}
\end{equation}
\end{widetext}
where $\Xi_\mu$ and $\Xi_{\mu \nu}$ denote the terms obtained from Eqs.~\eqref{EomemGV} and \eqref{hodeomA4}, respectively, after simply replacing the gravitational Lagrangian $\mathcal{L}_G$ by $\mathcal{L}_M$.

The equation of motion for $\Phi^{I}$ can also be obtained at this stage; it is produced by setting to zero the following quantity:
\begin{equation}
    \mathcal{E}_I = \frac{\partial \mathcal{L}_M}{\partial \Phi^I}- (-1)^{p} \D E_{I}^{(1)} + (-1)^{p(n-p)} \star G_I^{(0)}. \label{EOMMAtterPhi}
\end{equation}
Analogously, the equation of motion for $\mathit{A}^{I}_{\ J}$ is associated with
\begin{eqnarray}
 \mathcal{E}_{I}^{\ J} &=& \D [H^{(0)}]_{I}^{\ J} +  \star F_{\ L}^{J} \wedge [H^{(1)}]_{I}^{\ L} \nonumber \\
 &&- \star F_{\ I}^{L} \wedge [H^{(1)}]_{L}^{\ J} +  \star \Phi^{J} \wedge P_{I}^{(1)} + \Phi^{J} \wedge E_{I}^{(1)} \nonumber \\
 && + \sum_{i=1}^{h-1} \star \D^{i} \star F_{\ L}^{J} \wedge [H^{(i+1)}]_{I}^{\ L} \nonumber \\
 && - \sum_{i=1}^{h-1}\star \D^{i} \star F_{\ I}^{L} \wedge [H^{(i+1)}]_{L}^{\ J} \nonumber \\
 &&+ \sum_{i=1}^{k-1} \star \D^{i} \Phi^{J} \wedge E_{I}^{(i+1)} \nonumber \\
 &&+ \sum_{i=1}^{l-1} \star \D^{i} \star \Phi^{J} \wedge P_{I}^{(i+1)}.
 \label{EOMmatterA}
\end{eqnarray}
These are the objects needed to find the equations of motion for all the dynamical fields under consideration, which are given for completeness.

\section{Hamiltonian Charges} \label{ChargesAtInf}

To construct Hamiltonian charges in the used framework, it is necessary to formulate the Hamiltonian in terms of differential forms. A systematic procedure is presented in Ref.~\onlinecite{PhysRevD.103.064062}. However, for a general diffeomorphism-invariant theory, the approach developed in Ref.~\onlinecite{PhysRevD.72.104020} is more convenient, which is the one adopted throughout this work.

The Hamiltonian construction requires a decomposition of spacetime into space and time. To this end, let $\xi$ be a vector field generating time translations at infinity and assume that spacetime is globally hyperbolic, which admits a foliation of the form $\Sigma \times \mathbb{R}$, with $\Sigma$ a Cauchy surface. Let $\sigma_\xi : \Sigma \rightarrow M$ denote an embedding that maps points on $\Sigma$ into spacetime along the flow generated by $\xi$. With this structure, the Lagrangian can be written as~\cite{PhysRevD.103.064062}
\begin{equation}
    \mathcal{L} = e^0 \wedge \sigma_\xi^{*} \, \mathrm{i}_\xi \mathcal{L},
    \label{lagrangiansplit}
\end{equation}
where $e^0$ is chosen such that $\mathrm{i}_\xi e^0 = 1$ and $\sigma_\xi^{*}$ denotes the pullback induced by $\sigma_\xi$. This pullback restricts all fields to $\Sigma$ and it is understood that all fields are evaluated on $\Sigma$ via this projection.

Following the arguments of Ref.~\onlinecite{PhysRevD.72.104020}, the Hamiltonian density $\mathcal{H}[\xi]$ must be an $(n{-}1)$-form and can be constructed defining the ``velocities'' $(\mathcal{L}_\xi \phi)$ and their conjugate momenta $(\pi_\phi)$ as
\begin{equation}
    \mathcal{H}[\xi] = \mathcal{L}_\xi \phi \wedge \pi_\phi - \text{i}_\xi \mathcal{L}.
    \label{HamiltonianSch}
\end{equation}
For theories whose Lagrangian contains higher-order derivatives, these momenta are not given by the usual first–order definition. Instead, they arise from the boundary terms generated in the variation of the higher-derivative Lagrangian. This procedure is the Ostrogradsky method~\cite{Woodard2015Ostrogradsky,VAldaya_1980}, which 
provides the correct generalized canonical momenta appearing in 
$\mathcal{H}[\xi]$.

The Ostrogradsky method (including its generalization for field theories~\cite{VAldaya_1980}) consists of redefining the generalized velocities as new coordinate variables, $d^{i}q/dt^{i} = Q_{i}$, and associating to each $Q_{i}$ its conjugate momentum, defined as the quantity that appears in the boundary term together with the variation of $Q_{i}$. Importantly, this construction introduces a constraint on the resulting phase space.

An illustrative example is a Lagrangian of the form $L(q,\dot{q},\ddot{q})$. Its variation is given by
\begin{eqnarray}
    \delta L &=& \delta q \left[ 
        \frac{\partial L}{\partial q}
        - \frac{d}{dt} \left(  \frac{\partial L}{\partial \dot{q}} \right) 
        + \frac{d^{2}}{dt^{2}} \left( \frac{\partial L}{\partial \ddot{q}} \right) 
    \right] \nonumber \\
    && + \frac{d}{dt} \left[
        \delta q \left( 
            \frac{\partial L}{\partial \dot{q}} 
            - \frac{d}{dt} \frac{\partial L}{\partial \ddot{q}} 
        \right)
        + \delta \dot{q} \, \frac{\partial L}{\partial \ddot{q}}
    \right].
    \label{Lagrangianclassicalmechanics}
\end{eqnarray}
According to the Ostrogradsky construction one introduces new coordinate variables $Q_{1} = q$ and $Q_{2} = \dot{q}$, the latter of which constitutes a constraint between the original variables. The canonical momenta are then read directly from the boundary term in the variation, yielding 
\begin{eqnarray}
    P_{1} &=& \frac{\partial L}{\partial \dot{q}} 
             - \frac{d}{dt} \left( \frac{\partial L}{\partial \ddot{q}} \right), \\
    P_{2} &=& \frac{\partial L}{\partial \ddot{q}} .
\end{eqnarray}
By definition, the Hamiltonian is constructed as
\begin{equation}
    H = \dot{Q_{1}} P_{1} + \dot{Q_{2}} P_{2} - L,
    \label{HamiltonianCM}
\end{equation}
which is understood as a functional of the canonical variables $(Q_{1}, Q_{2}, P_{1}, P_{2})$. 

In the gravitational case, the curvature and torsion 2-forms contain temporal derivatives of the spin connection and veilbein, respectively. From the structure of the boundary term it can be seen that
\begin{eqnarray}
 \theta_G(\delta) &=& \delta \omega^{\mu \nu} \wedge H_{\mu \nu}^{(0)}+ \delta e^\mu \wedge E_{\mu}^{(0)} \nonumber \\
 &&+ \delta \star R^{\mu \nu} \wedge H_{\mu \nu}^{(1)}+ \delta \star T^{\mu} \wedge E_{\mu}^{(1)} \nonumber \\
 && +\sum_{i=1}^{N-1} \delta \star \D^{i} \star R^{\mu \nu} \wedge H_{\mu \nu}^{(i+1)} \nonumber \\
 &&+ \sum_{i=1}^{M-1}\delta \star \D^{i} \star T^{\mu} \wedge E_{\mu}^{(i+1)}.
 \label{BTG}
\end{eqnarray}
Defining $\chi^{(i)\mu\nu} = \star \D^{i} \star R^{\mu\nu}$ and $\psi^{(i)\mu} = \star \D^{i} \star T^{\mu}$, it is evident that these variables encode information about higher-order time derivatives. Following the Ostrogradsky method, the canonical momenta associated with $\omega^{\mu\nu}$ and $e^{\mu}$ are
\begin{eqnarray}
    \pi^{(e)}_{\mu} &=& E_{\mu}^{(0)}, \\
    \pi^{(\omega)}_{\mu\nu} &=& H_{\mu\nu}^{(0)} ,
    \label{Momentums}
\end{eqnarray}
where these quantities can be read off directly from the corresponding boundary term. Proceeding recursively, the canonical momenta conjugate to $\chi^{(i)\mu\nu}$ and $\psi^{(i)\mu}$ are obtained by identifying the coefficients of their variations in the boundary term.

With these identifications, it can be verified that
\begin{eqnarray}
 \theta_{G}(\mathcal{L}_{\xi}) &=& 
   \mathcal{L}_{\xi}\omega^{\mu\nu} \wedge \pi^{(\omega)}_{\mu\nu}
 + \mathcal{L}_{\xi}e^{\mu}      \wedge \pi^{(e)}_{\mu} \nonumber \\
 &&+\, \mathcal{L}_{\xi}\star R^{\mu\nu} \wedge \pi^{(1)}_{\mu\nu}
   + \mathcal{L}_{\xi}\star T^{\mu}    \wedge \pi^{(1)}_{\mu} \nonumber \\
 &&+\, \sum_{i=1}^{n-1}
        \mathcal{L}_{\xi}\star\D^{i}\star R^{\mu\nu} \wedge \pi^{(i)}_{\mu\nu}
    \nonumber \\
 &&+\, \sum_{i=1}^{m-1}
        \mathcal{L}_{\xi}\star\D^{i}\star T^{\mu} \wedge \pi^{(i)}_{\mu}
    \nonumber \\
 &=& \mathcal{L}_{\xi}\phi \,\wedge\, \pi_{\phi},
 \label{BTGLie}
\end{eqnarray}
where $\mathcal{L}_{\xi}\phi$ denotes the ``velocities'' of the dynamical fields, while $\pi_{\mu}^{(i)}$ and $\pi_{\mu\nu}^{(i)}$ represent their corresponding canonical momenta. In this sense, the Hamiltonian construction obtained by combining the Ostrogradsky and Nester methods~\cite{CHEN19955} yields a Hamiltonian functional that is expected to encode the physical charges at spatial infinity, although its validity may require additional assumptions in nonstandard higher-derivative theories.

Using Eq.~\eqref{Ltexti}, $H[\xi]$ takes the form
\begin{eqnarray}
    H[\xi] = \int_\Sigma \mathcal{H}[\xi] 
    &=& \int_\Sigma \big[ -\,\text{i}_{\xi} e^{\mu}(\mathcal{E}_\mu - \tau_{\mu}) \nonumber \\
    && \ \ \ - \text{i}_{\xi} \omega^{\mu \nu} (\mathcal{E}_{\mu \nu} - S_{\mu \nu}) \nonumber \\
    && \quad -\, \text{i}_{\xi} \Phi^{I} \wedge \mathcal{E}_{I} 
    - \text{i}_{\xi} \mathit{A}^{I}_{\ J} \mathcal{E}_{I}^{\ J} 
    + \de Q(\xi) \big], \nonumber \\
    \label{HamiltonianFunctional}
\end{eqnarray}
where $H[\xi]$ is projected onto $\Sigma$, and the equations of motion correspond to the constraint equations~\cite{PhysRevD.103.064062}, and 
\begin{equation}
\mathcal{H}[ \xi ] = \theta(\mathcal{L}_\xi) - \text{i}_\xi \mathcal{L} = \mathcal{L}_\xi \phi \wedge \pi_\phi - \text{i}_\xi \mathcal{L}. 
\label{HamiltonianAprox}
\end{equation}
Moreover, Eq.~\eqref{HamiltonianFunctional} exhibits the general structure of the Hamiltonian in diffeomorphism-invariant theories~\cite{WaldB}. In this expression, $Q(\xi)$ represents the Noether charge associated with diffeomorphisms, which also appears as the boundary term in the Hamiltonian.

As is well known~\cite{WaldB,MR359663}, the variation of the Hamiltonian may produce additional boundary contributions. These terms must be adjusted so that the condition in Eq.~\eqref{HamiltonianCI} holds on shell. Using Eq.~\eqref{HamiltonianAprox}, the variation of the Hamiltonian takes the form~\cite{CHEN19955,PhysRevD.72.104020}
\begin{equation}
    \delta H[\xi] = \Omega (\delta, \mathcal{L}_\xi) + \int_\Sigma \de \text{i}_\xi \theta (\delta).
    \label{varH}
\end{equation}
For an asymptotically flat configuration, and if the presymplectic current satisfies $\theta(\delta) = \delta B$, where $B$ is a local functional of the dynamical fields~\cite{Iyer}, then the modified Hamiltonian $\bar{H}[\xi]$ that satisfies Eq.~\eqref{HamiltonianCI} can be written as
\begin{equation}
    \bar{H}[\xi] = H[\xi] - \int_\Sigma \de \text{i}_\xi B,
    \label{functionalH}
\end{equation}
or, equivalently,
\begin{equation}
    \bar{H}[\xi] = \int_\Sigma \mathcal{C}[\phi] + \de \big[ Q(\xi) - \text{i}_\xi B \big],
    \label{FunctionalH2}
\end{equation}
where $\mathcal{C}[\phi]$ collects the constraint.

When the constraints are satisfied, and applying Stokes' theorem, the conserved charge at spatial infinity is given by
\begin{equation}
    \bar{H}[\xi] = \int_{\text{i}^0} Q(\xi) - \text{i}_\xi B ,
    \label{ChargeHamiltonian}
\end{equation}
and its variation satisfies
\begin{equation}
    \delta \bar{H}[\xi] = \int_{\text{i}^0} \delta Q(\xi) - \text{i}_\xi \theta(\delta),
    \label{varChargeHamiltonian}
\end{equation}
which coincides with the term appearing in the first law of black hole thermodynamics derived from the presymplectic current. 

What happens to the Hamiltonian and the associated charges at infinity when an exact form is added to the Lagrangian? As shown in Ref.~\cite{Iyer}, if $\mathcal{L} \rightarrow \mathcal{L}' = \mathcal{L} + \de \mu$, then the Noether charge and the presymplectic current transform as 
$Q(\xi) \rightarrow Q'(\xi) = Q(\xi) + \text{i}_\xi \mu$ and 
$\theta(\delta) \rightarrow \theta'(\delta) = \theta(\delta) + \delta \mu$, respectively. In this case, the Hamiltonian becomes
\begin{eqnarray}
    H'[\xi] &=& \int_\Sigma \theta'(\mathcal{L_\xi}) - \text{i}_\xi \mathcal{L}' \nonumber \\
    &=& H[\xi] + \int_\Sigma \de \text{i}_\xi \mu \nonumber \\
    &=& \int_\Sigma \mathcal{C}[\phi] + \de Q'(\xi),
    \label{HamiltonianBT}
\end{eqnarray}
and the corrected Hamiltonian takes the form
\begin{equation}
    \bar{H}'[\xi] = \int_\Sigma \mathcal{C}[\phi] + \de \big[ Q'(\xi) - \text{i}_\xi B' \big],
    \label{HamiltonianBT2}
\end{equation}
where $B'$ satisfies $\delta B' = \theta'(\delta)$. Thus, when the fields satisfy the constraints, the charges at infinity are
\begin{equation}
    \bar{H}'[\xi] = \int_{\text{i}^0} \big(Q'(\xi) - \text{i}_\xi B'\big).
    \label{HamiltonianBT3}
\end{equation}
This behavior is expected since adding boundary terms can modify the asymptotic properties of the theory. This quantity can be fixed to zero for the Minkoski spacetime (when it is a solution of the theory). Still, the variation of Eq.~\eqref{HamiltonianBT3} yields the same presymplectic current, implying that the physical content of the first law of black hole thermodynamics is invariant under $\mathcal{L} \rightarrow \mathcal{L}' = \mathcal{L} + \de \mu$.

\bibliography{References}
\end{document}